\documentclass[journal]{IEEEtran}
\usepackage{amsmath,amsfonts}
\usepackage{algorithmic}
\usepackage{algorithm}
\usepackage{array}
\usepackage[caption=false,font=normalsize,labelfont=sf,textfont=sf]{subfig}
\usepackage{textcomp}
\usepackage{stfloats}
\usepackage{url}
\usepackage{verbatim}
\usepackage{graphicx}
\usepackage{booktabs}
\usepackage{cite}
\usepackage{tabularx,booktabs}
\usepackage{diagbox}
\usepackage{array,multirow}
\usepackage{wrapfig}
\usepackage{lipsum}
\usepackage{mathtools}
\usepackage{float}
\usepackage{xcolor}
\usepackage{enumitem}
\usepackage{csquotes}
\usepackage{makecell}
\usepackage{amssymb}
\usepackage{booktabs}
\usepackage{adjustbox}
\usepackage{colortbl}
\usepackage{longtable}
\usepackage{lscape}
\usepackage{tabularx}
\usepackage{colortbl,booktabs}
\usepackage[%
	acronym=true,%
	nonumberlist,%
	nopostdot,%
	seeautonumberlist,%
	shortcuts,%
	section=chapter,%
	toc=false,%
	nogroupskip=true,%
    prefix%
]{glossaries-extra}
\usepackage{textcase,mfirstuc,xspace}
\usepackage{orcidlink}
\usepackage{comment}
\usepackage{svg}
\usepackage{soul}
\usepackage{pifont}
\usepackage[most]{tcolorbox}
\usepackage{eso-pic}
\newcolumntype{C}{>{\centering\arraybackslash}X}
\usepackage[spaces,hyphens]{xurl} % 
\usepackage{booktabs,xcolor,siunitx}
\definecolor{headergray}{RGB}{162, 182, 215}    
\definecolor{maincolor}{RGB}{55, 90, 140}       
\definecolor{subcolor}{RGB}{90, 110, 140}      
\definecolor{lightgray}{RGB}{255, 204, 102}       
\usepackage{hyperref}

\renewcommand{\eqref}[1]{\textup{\textcolor{blue}{(}\ref{#1}\textcolor{blue}{)}}}
\hypersetup{%
  colorlinks=true,%
  linkcolor={blue!65!black},
  citecolor={blue},
  urlcolor={blue!80!black},
  bookmarksnumbered=true,%
  bookmarksopen=true}
\begin{document}   
\title{A Comprehensive Survey of Channel Estimation Techniques for OTFS in 6G and Beyond Wireless Networks}
\author{Emir Aslandogan,~Haci Ilhan,~\IEEEmembership{Senior Member,~IEEE},~Burak Ahmet Ozden,~Erdogan Aydin,~Ertugrul Basar,~\IEEEmembership{Fellow,~IEEE},~Miaowen Wen, ~\IEEEmembership{Senior Member,~IEEE},~Marco Di Renzo,~\IEEEmembership{Fellow,~IEEE},~H. Vincent Poor,~\IEEEmembership{Life Fellow,~IEEE}
\thanks{E. Aslandogan and H. Ilhan are with Y{\i}ld{\i}z Technical University, Department of Electronics and Communications Engineering, 34220, Davutpasa, Istanbul, Turkey (e-mail: emira@yildiz.edu.tr and ilhanh@yildiz.edu.tr).}  
\thanks{B. A. Ozden is with the Department of Electrical and Electronics
Engineering, Istanbul Medeniyet University, 34857 Istanbul, Türkiye, and also
with the Department of Computer Engineering, Yıldız Technical University,
34220 Istanbul, Türkiye (e-mail: bozden@yildiz.edu.tr).}
\thanks{E. Aydin is with the Department of Electrical and Electronics Engineering, Istanbul Medeniyet University, 34857 Istanbul, Türkiye (e-mail:
erdogan.aydin@medeniyet.edu.tr)}
\thanks{E. Basar is with the Department of Electrical Engineering,
Tampere University, 33720 Tampere, Finland, on leave from the Department
of Electrical and Electronics Engineering, Koc University, 34450 Sariyer,
Istanbul, Turkey (e-mail: ertugrul.basar@tuni.fi).}
\thanks{M. Wen is with School of Information Science and Technology, Nantong
University, Nantong 226019, China, and also with School of Electronic and
Information Engineering, South China University of Technology, Guangzhou
510640, China (e-mail: eemwwen@scut.edu.cn). }
\thanks{M. Di Renzo is with Universit\'e Paris-Saclay, CNRS, CentraleSup\'elec, Laboratoire des Signaux et Syst\`emes, 3 Rue Joliot-Curie, 91192 Gif-sur-Yvette, France. (marco.di-renzo@universite-paris-saclay.fr), and with King's College London, Centre for Telecommunications Research -- Department of Engineering, WC2R 2LS London, United Kingdom (marco.di\_renzo@kcl.ac.uk).}
\thanks{H. V. Poor is with the Department of Electrical and Computer Engineering, Princeton University, Princeton, NJ 08544 USA (e-mail: poor@princeton.edu).}}
%\markboth{IEEE Communications Surveys \& Tutorials}
\maketitle
\begin{abstract}
Orthogonal time-frequency space (OTFS) modulation has emerged as a powerful wireless communication technology that is specifically designed to address the challenges of high-mobility scenarios and significant Doppler effects.  Unlike conventional modulation schemes that operate in the time-frequency (TF) domain, OTFS projects signals to the delay-Doppler (DD) domain, where wireless channels exhibit sparse and quasi-static characteristics. This fundamental transformation enables superior channel estimation (CE) performance in challenging propagation environments characterized by high-mobility, severe multipath effects, and rapidly time-varying channel conditions. This article provides a systematic examination of CE techniques for OTFS systems, covering the extensive research landscape from foundational methods to cutting-edge approaches. We present a detailed analysis of DD and TF domain CE techniques presented in the literature, including separate pilot, embedded pilot, and superimposed pilot approaches. The article encompasses various algorithmic frameworks including Bayesian learning, matching pursuit-based techniques, message passing algorithms, deep learning (DL)-based methods, and recent CE approaches. Additionally, we explore joint CE and signal detection (SD) strategies, the integration of OTFS with next-generation wireless systems including massive multiple-input multiple-output (MIMO), millimeter wave (mmWave) communications, reconfigurable intelligent surfaces (RISs), and integrated sensing and communication (ISAC) systems. Critical implementation challenges are presented, including leakage suppression, inter-Doppler interference mitigation, impulsive noise handling, signaling overhead reduction, guard space requirements, peak-to-average power ratio (PAPR) management, beam squint effects, and hardware impairments. 
\end{abstract}
\maketitle
\section{Introduction}
\label{sec:introduction}
\IEEEPARstart{T}{he} continuing development of sixth generation (6G) and beyond wireless networks is creating increasingly high-mobility system scenarios.  Beyond fifth generation (5G), these emerging scenarios are characterized by increasing complexity. It requires the development of more robust wireless communication technologies \cite{8808168,8869705}. Applications involving low-Earth-orbit (LEO) satellites \cite{9804811}, high-speed trains, and unmanned aerial vehicle (UAV) networks \cite{7470933} are primary examples that introduce such high-mobility channel conditions. As a result, traditional orthogonal frequency division multiplexing (OFDM), despite its widespread use, experiences significant performance degradation in these challenging propagation environments \cite{10794219}.

\begin{table*}[!htbp]
\centering
\caption{List of Abbreviations Used in the Article}
\renewcommand{\arraystretch}{0.8}
\scriptsize
\begin{tabular}{|c|p{6cm}|c|p{6cm}|}
\hline
\rowcolor{lightgray}
\textbf{Abbreviation} & \textbf{Description} & \textbf{Abbreviation} & \textbf{Description} \\
\hline \hline
2D & two-dimensional & LEn-CENet & lightweight enhanced CE network \\
3D-ESP & 3d efficient subspace pursuit & LS & least-squares \\
3D-IPRDSOMP & 3d inner product proportion reduce difference structured OMP & LSTM & long short-term memory \\
3D-SOMP & 3d-structured OMP & M-CPSBL & multiple-frame coupled prior SBL \\
5G & fifth generation & MDCFT & modified discrete chirp Fourier transform \\
6G & sixth generation & MEP & message passing \\
ABOMP & adaptive block OMP & MF-OAMEP & mean-field orthogonal approximate message passing \\
ADC & analog-to-digital converter & MIMO & multiple-input multiple-output \\
AP-SP & affine-precoded superimposed pilot & MLS & maximum length sequence \\
ASGR & adaptive second-order grid refinement & MM & majorization-minimization \\
AWGN & additive white Gaussian noise & mmWave & millimeter wave \\
BER & bit error rate & MMSE & minimum mean square error \\
BL & Bayesian learning & MP & matching pursuit \\
BR & block reorganization & MSP & modified subspace pursuit \\
BSBL & block sparse Bayesian learning & MSSAMP & modified SSAMP \\
CD-SIC & cross-domain serial interference cancellation & MVC & maximum Versoria criterion \\
CE & channel estimation & NOMA & non-orthogonal multiple access \\
CIPP & corner-inserted pilot pattern & OFDM & orthogonal frequency division multiplexing \\
CNN & convolutional neural network & OG-BCS & off-grid Bayesian compressive sensing \\
COMP & combined OMP & OMP & orthogonal matching pursuit \\
CP & cyclic prefix & OSBL & online sparse Bayesian learning \\
CRLB & Cramér–Rao lower bound & OTFS & orthogonal time-frequency space \\
CS & compressed sensing & PACESD & parameter association, CE and SD \\
CS-BEM & CS-basis expansion model & PAPR & peak-to-average power ratio \\
CSC & convolutional sparse coding & PBiGaBP & parametric bilinear Gaussian belief propagation \\
DD & delay-Doppler & PI-SBL & parameter-inherited SBL \\
DDA & delay-Doppler-angle & PN & pseudo-noise \\
DDIPIC & DD inter-path interference cancellation & PSO & particle swarm optimization \\
DFT & discrete Fourier transform & RBOMP & row-block orthogonal matching pursuit \\
DL & deep learning & ResNet & residual network \\
DNN & deep neural network & RG & row-group \\
DRAN & deep residual attention network & RIS & reconfigurable intelligent surface \\
DRSNN & deep residual shrinkage neural network & RNN & recurrent neural network \\
DT & delay-time & S-SBL & segmentation-based SBL \\
DZT & discrete Zak transform & SBL-JE & SBL-based joint channel and clipping amplitude estimator \\
EKF & extended Kalman filter & SCMA & sparse code multiple access \\
EM & expectation maximization & SD & signal detection \\
EM-VBL & EM-based VBL & SDR & semi-definite relaxation \\
EP & embedded pilot & SER & symbol error rate \\
ESBL & efficient SBL & SL0 & smoothed l0 \\
FBCS & fast variant BCS & SNR & signal-to-noise-ratio \\
FD-PA-OTFS & frequency-domain pilot-aided OTFS & SP & superimposed pilot \\
FM-CPSBL & fast multiple-frame-based-CPSBL & SPA & sum-product algorithm \\
GAMEP & generalized approximate message passing & SSAMP & structured sparsity adaptive MP \\
GAN & generative adversarial network & SSIP & superimposed single pilot \\
HPA & high-power amplifier & SSPP & superimposed sparse pilot \\
HRIS & hybrid RIS & ST & single-tone \\
HSP & hybrid superimposed pilot & T-GEESBI & truncated grid evolution-based efficient sparse Bayesian inference \\
ICI & inter-carrier interference & TD-PA-OTFS & time-domain pilot-aided OTFS \\
IDI & inter-Doppler interference & TD-TS-OTFS & time-domain training sequence-based OTFS \\
IGI & inter-grid interference & TF & time-frequency \\
IIC & iterative interference cancellation & THz & terahertz \\
IPPS & iterative path peak search & TSICE & two-stage iterative channel estimation \\
ISAC & integrated sensing and communication & UAV & unmanned aerial vehicle \\
ISFFT & inverse symplectic finite Fourier transform & UAMEP & unitary approximate message passing \\
ISI & inter symbol interference & UWA & underwater acoustic \\
JCEDD & joint CE, data detection, and demodulation & V2X & vehicle-to-everything \\
L-BFGS & limited-memory Broyden–Fletcher–Goldfarb–Shanno & VBI & variational Bayesian inference \\
LEO & low-earth-orbit & VBL & variational BL \\
 & & WLS & weighted LS \\
 & & ZBLP & zero-bin low-papr \\
 & & ZC & Zadoff-Chu \\
\hline
\end{tabular}
\label{tab:abbr}
\end{table*}

The development of orthogonal time-frequency space (OTFS) modulation is particularly crucial for dynamic system scenarios where the Doppler effect is strong \cite{hadani2017orthogonal}. Therefore, accurate channel estimation (CE) is essential for ensuring reliable communication performance in doubly-dispersive channels. It operates in the OTFS delay-Doppler (DD) domain instead of the traditional time-frequency (TF) domain. This provides a different framework for channel characterization and is advantageous in environments where traditional modulation schemes demonstrate performance limitations \cite{aslandogan2025biterrorrateperformance}.

In OTFS systems, the DD domain demonstrates its benefits, especially in terms of CE. Sparse and semi-static channel representation reduces the scale of the estimation problem \cite{Hong_2022}. This sparsity is not an artifact of the
transformation but rather reveals the true physical nature of
wireless propagation, where only a limited number of reflectors
contribute to the received signal. The inherent characteristics of the DD domain have resulted in an evolution of advanced estimation methodologies that leverage this sparsity. This decreases pilot overhead and computational complexity while improving precision. These developments have increased interest in OTFS technology and accelerated research into developing efficient and robust CE techniques in this domain. These efforts include a wide variety of approaches, ranging from traditional pilot methods to advanced deep learning (DL) algorithms. Each method has its own advantages and disadvantages in terms of accuracy, complexity, and applicability. The wide range of solutions reflects the importance of OTFS-based CE and the complexities of the associated techniques.

\subsection{OTFS Application Scenarios}
\label{subsec:otfs_wireless_tech}
The consideration of OTFS modulation for various wireless communication systems has been motivated by its resilience in dynamic environments. In particular, it is an appropriate technology for new applications that need reliable performance under challenging propagation conditions because it provides a stable channel representation in high-mobility scenarios.

\begin{figure*}
    \centering
    \includegraphics[width=.75\linewidth]{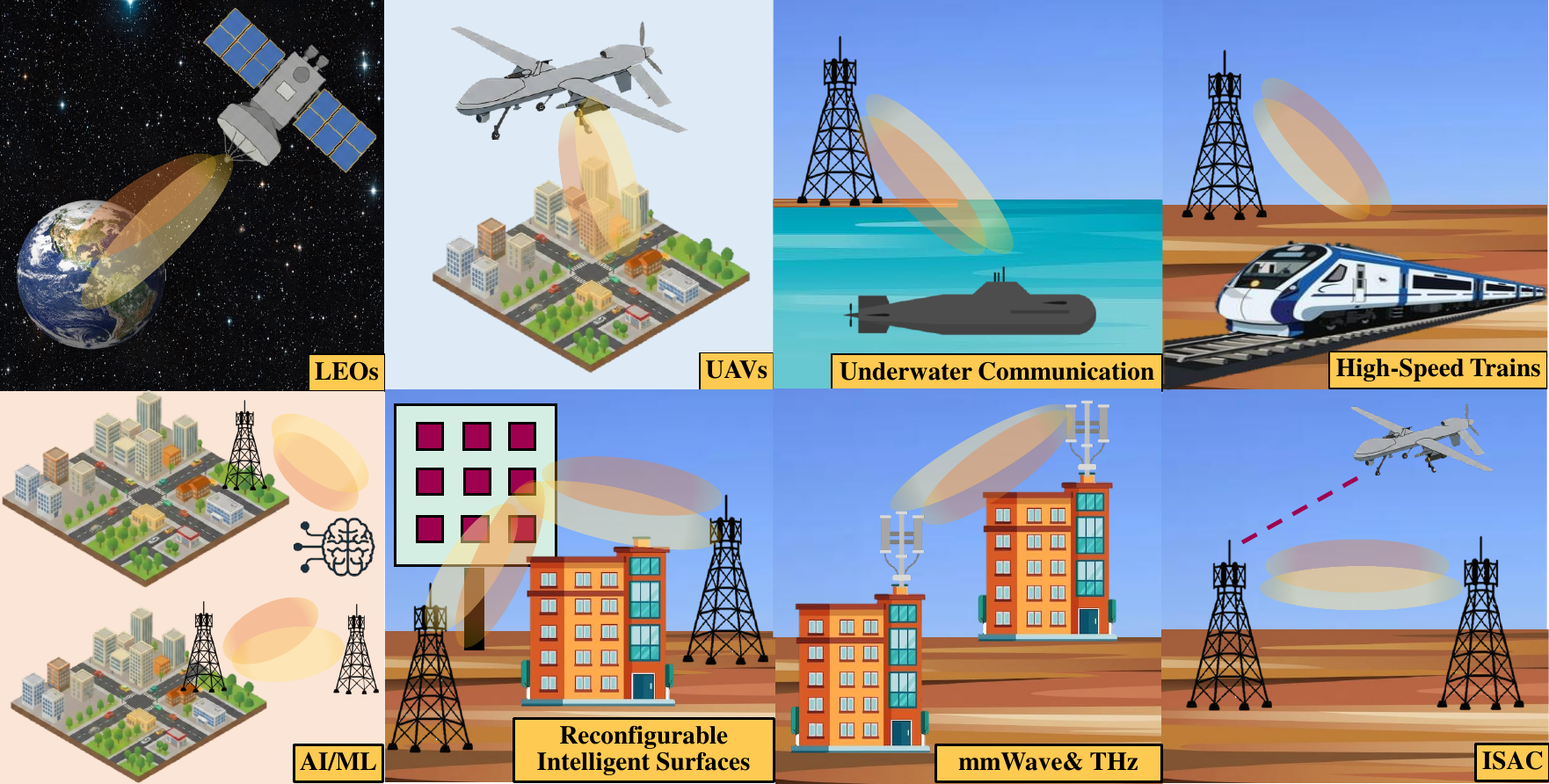}
    \caption{Applications of OTFS in wireless communication systems.}
    \label{fig:Wireless}
\end{figure*}

As illustrated in Fig. \ref{fig:Wireless}, the characteristics of OTFS have led to its investigation in many important application areas. For example, in integrated sensing and communication (ISAC), OTFS's DD domain representation provides a unified structure for both radar sensing and communication signals. Studies focus on OTFS's resilience to severe Doppler effects for networks with high mobility, including UAVs and LEO satellites. With the aim to enhance spectral efficiency, current research includes investigating into integrating OTFS with reconfigurable intelligent surfaces (RISs). To mitigate inter-carrier interference (ICI), applications in the terahertz (THz) and millimetre wave (mmWave) bands are also being investigated. In underwater acoustic (UWA) communications, OTFS has been investigated to counteract the severe multipath and Doppler effects characteristic of aquatic environments. Across many of these applications, DL techniques are also being widely investigated as a tool to improve the performance of CE and signal detection (SD) in OTFS systems. In each of these domains, OTFS is studied as an alternative to conventional modulation, offering potential solutions to distinct propagation challenges.

\subsubsection{ISAC}
The purpose of the rapidly evolving area of ISAC research is to integrate communication and sensing capabilities into a single system framework. ISAC provides various advantages over traditional separated designs, such as lower costs and increased spectral efficiency, by allowing resource sharing across hardware and spectrum \cite{9296833,9585321}. Due to its operational characteristics in the DD domain, OTFS has been identified as a promising waveform for these dual-function applications \cite{7855671, 8999605}. Some initial investigations into ISAC considered OTFS-ISAC configurations in attempts to unify radar sensing and communication functions \cite{Gaudio_2020}. As OTFS operates in the DD domain, enabling the estimation of the channel's Doppler and delay parameters, it is well-suited to ISAC.  There is a natural alignment between the two tasks because these parameters are similar to the target range and velocity information needed in radar. Because of this property, many studies have examined OTFS-based joint radar-communication systems \cite{10409528,10318060,10714398,11058951,10770206,10309871,10845803,11066286,10265224,10251770,10192916}.

\subsubsection{Deep Learning}
DL is used as a tool to solve signal processing challenges in OTFS systems. It is challenging to model and optimize the DD domain in OTFS using conventional techniques for CE, detection, and resource allocation due to several distinctive features. DL algorithms offer a potentially promising data-driven approach to improve system performance by learning complex patterns directly from the data. Research in this area mainly concentrated on CE \cite{9940260,channelEstimationReff,11071963,10654761,10439989,10856399,10138432,10916513, 10747184,10012974} and SD \cite{9448630,10018844,10142162,symbolDetectionReff,10154051,10978604,9814608, enku2021two,10105487}, and additional studies exploring DL applications in broader system architectures and frameworks \cite{10138552,9569634,kumar2023residual,10226266,10375127}.

\subsubsection{UAVs}
OTFS modulation has been studied as a potential technology for UAV communications, essentially due to its resilience in highly dynamic environments. OTFS's resilience to high Doppler shifts makes this technology particularly suitable for establishing reliable connections for UAV control and data transmission in dynamic wireless channels. In this context, a key advantage of OTFS is its ability to provide direct estimates of target range and velocity in the DD domain, which is its natural capability for joint communication and sensing. In particular, OTFS can resolve paths with the same delay but different Doppler shifts, which is a challenging task for traditional OFDM and is crucial for accurate velocity estimation in multipath scenarios \cite{10891196}. Consequently, studies on applying OTFS to UAVs focus on high-precision localization \cite {10437569, 9685862}, positioning \cite{10663682, 10038846}, and other topics \cite{11008363,10978356,10979952,10891196}.

\subsubsection{LEOs}
LEO satellite constellations provide worldwide coverage; however, they are characterized by significant Doppler shifts with rapid channel variations as a result of their high velocities. These conditions require resilient waveforms, leading to OTFS, an appropriate choice for dependable LEO communication links. Extensive works on these challenges. The literature spans a wide range of domains, from CE and equalization \cite{9785832}, MIMO-OTFS \cite{9814666,9849120,10680158}, random access \cite{9849120}, and cooperative transmission \cite{10679987,10680158} to critical performance metrics like outage probability \cite{9689960}, secrecy \cite{10465059}, and detection \cite{10437523}.

\subsubsection{RISs}
RIS-OTFS integration, which stands out as a solution to the challenges faced by next-generation wireless networks, is a growing area of research. This method aims to optimize system performance by combining the beamforming capabilities of RIS technology with the inherent robustness advantage of OTFS in high-mobility scenarios. Current research in RIS-OTFS spans a variety of topics. Key areas of investigation include the enhancement of high-Doppler channels \cite{RISOTFS27, RISOTFS25}, index modulation (IM) techniques \cite{OTFS_RIS1,OTFS_IM11,OTFS_IM12,ozden2025surveyotfsbasedindexmodulation}, wavelet-based implementations \cite{RISOTFS16}, and passive beamforming \cite{RISOTFS12}. Additionally, specific use cases such as localization and positioning \cite{RISOTFS23,11079715,10017173}, high-speed railways \cite{10979982}, space–air–ground integrated networks (SAGINs) \cite{RISOTFS1}, and optical communications \cite{10676782} have been the subject of recent investigations.

\subsubsection{mmWave \& THz Communication}
The studies of mmWave and THz frequency bands are motivated by the necessity to meet the high data rate requirements of future wireless networks; yet, their vulnerability to performance degradation caused by Doppler effects poses a significant challenge for conventional OFDM systems. Consequently, OTFS is investigated as a more robust alternative for high-mobility mmWave scenarios, as its operation in the DD domain can effectively mitigate ICI \cite{9417346,9827947,10100890}. In the THz band, in addition to severe Doppler effects, system designs must also contend with strict peak-to-average power ratio (PAPR) requirements. To overcome these limitations, methods such as discrete Fourier transform (DFT)-spread OTFS have been proposed, mainly to improve high-capacity communications and high-resolution sensing \cite{10061469}. Consequently, the existing literature offers a broad spectrum of research that presents novel frameworks and signal processing methodologies tailored for OTFS-based mmWave and THz communication systems \cite{8746382,10597075,10100890,10978739,9559057,9814573,10918701,8580850,11069272,11084839,10857395,10916590}.

\subsubsection{Underwater Communication}
UWA channels, considered one of the most challenging propagation environments, are characterized by wide delay spreads, severe Doppler effects, and limited bandwidth, which is further amplified by the dynamic water environment. OTFS modulation is being investigated for UWA systems because it transforms this complex, time-varying channel into a sparse representation in the DD domain. This property is advantageous for achieving reliable performance in such doubly-dispersive channels \cite{li2024underwater, xue2023orthogonal}. To address the signal processing difficulties in UWA environments, extensive studies have been conducted on CE and tracking \cite{OTFSUnderWater19}, pilot design \cite{OTFSUnderWater7}, SD and receiver architectures \cite{10400942,10423132,OTFSUnderWater17,10870146}, as well as adaptive modulation and coding schemes \cite{10569021}.
\subsection{Contributions of This Survey}
In recent years, while various surveys have covered the fundamentals of OTFS and CE, there has been a need for research focusing on OTFS-CE techniques and the algorithms used. This article aims to provide a structured overview and classification of OTFS-CE approaches to support ongoing developments in this direction. The contributions of this article are as follows:
\begin{itemize}
\item This article aims to outline a framework for approaches in the literature by classifying OTFS-CE techniques, primarily DD-domain approaches (separate pilot, embedded pilot, and superimposed pilot), TF-domain methods, and algorithm-based approaches such as Bayesian learning (BL), matching pursuit (MP), message passing (MEP), and DL-based methods. 
\item This article examines at over 200 studies, from foundational works to advances by mid-2025.  We examine research addressing fundamental application challenges including leakage suppression, mitigation of inter-Doppler interference (IDI), impulsive noise management, reduction of signal overhead, guard space requirements, PAPR control, beam squint effects, and hardware impairments. 
\item Within the framework of next-generation wireless systems, this study examines OTFS-CE, highlighting the key difficulties and possible avenues for future research in areas such as massive MIMO, mmWave communications, RIS, ISAC, UAV communications, LEO satellite systems, and UWA communications.
\item This article summarizes findings from earlier studies. Performance comparison under various channel conditions, parameter optimization techniques, and complexity analysis are covered. This gives professionals an insight into what to consider when deploying an OTFS system in the real world.
\item This article highlights significant open problems. It outlines future research directions based on the chronological progress of OTFS-CE. The article reviews recent developments to spot new trends. In addition, it offers insight into potential paths for future work.
\end{itemize}
The paper is organized as shown in Table~\ref{tab:survey_roadmap}. The OTFS system model provides the framework in Section~\ref{sec:background}. Section~\ref{sec:CH_ES} discusses DD domain methods, while Section~\ref{sec:tf_domain} discusses TF domain approaches. Algorithmic solutions, including BL, MP, MEP, and DL, are reviewed in Section~\ref{sec:ce_algorithms}. Furthermore, Section~\ref{sec:joint_estimation} examines joint CE and SD, while Section~\ref{sec:next_gen_systems} addresses integration with next-generation systems. Finally, Section~\ref{sec:OTFS_Challenges} highlights key challenges for practical deployment, such as leakage suppression and interference mitigation.

\begin{table}[!htbp]
\centering
\caption{Survey Organization and Structure}
\renewcommand{\arraystretch}{0.855}
\label{tab:survey_roadmap}
\begin{tabular}{|p{0.95\columnwidth}|}
\hline
\rowcolor{lightgray}
\textbf{Survey Structure} \\
\hline
\quad \hyperref[sec:background]{\textbf{Section II: Background of OTFS Modulation}} \\[2pt]
\quad\quad $\triangleright$ \hyperref[subsec:transmitter]{Transmitter} \\[1pt]
\quad\quad $\triangleright$ \hyperref[subsec:channel]{Channel} \\[1pt]
\quad\quad $\triangleright$ \hyperref[subsec:receiver]{Receiver} \\[1pt]
\quad \hyperref[sec:CH_ES]{\textbf{Section III: OTFS Channel Estimation Techniques}} \\[2pt]
\quad\quad $\triangleright$ \hyperref[subsec:separate_pilot]{Separate Pilot Techniques} \\[1pt]
\quad\quad\quad $\circ$ \hyperref[subsubsec:impulse_based]{Impulse-based Channel Estimation} \\[1pt]
\quad\quad\quad $\circ$ \hyperref[subsubsec:pseudo_noise]{Pseudo-Noise Based Channel Estimation} \\[1pt]
\quad\quad $\triangleright$ \hyperref[subsec:embedded_pilot]{Embedded Pilot Techniques} \\[1pt]
\quad\quad\quad $\circ$ \hyperref[subsubsec:embedded_estimation]{Embedded Pilot Based Estimation} \\[1pt]
\quad\quad\quad $\circ$ \hyperref[subsubsec:sparse_bayesian]{Sparse Bayesian Learning Based Estimation} \\[1pt]
\quad\quad $\triangleright$ \hyperref[subsec:superimposed_pilot]{Superimposed Pilot Based Estimation Techniques} \\[2pt]
\quad \hyperref[sec:tf_domain]{\textbf{Section IV: Time-Frequency Domain Channel Estimation}} \\[2pt]
\quad\quad $\triangleright$ \hyperref[subsec:delay_estimation]{Delay Estimation} \\[1pt]
\quad\quad $\triangleright$ \hyperref[subsec:doppler_estimation]{Doppler Estimation} \\[1pt]
\quad\quad $\triangleright$ \hyperref[subsec:channel_gain_estimation]{Channel Gain Estimation} \\[2pt]
\quad \hyperref[sec:ce_algorithms]{\textbf{Section V: Signal Processing Algorithms for Channel Estimation}} \\[2pt]
\quad\quad $\triangleright$ \hyperref[subsec:bayesian_learning]{Bayesian Learning} \\[1pt]
\quad\quad $\triangleright$ \hyperref[subsec:matching_pursuit]{Matching Pursuit-Based Techniques} \\[1pt]
\quad\quad $\triangleright$ \hyperref[subsec:message_passing]{Message Passing-Based Techniques} \\[1pt]
\quad\quad $\triangleright$ \hyperref[subsec:deep_learning]{Deep Learning-Based Techniques} \\[1pt]
\quad\quad $\triangleright$ \hyperref[subsec:recent_techniques]{Recent Techniques} \\[2pt]
\quad \hyperref[sec:joint_estimation]{\textbf{Section VI: Joint Channel Estimation and Signal Detection}} \\[2pt]
\quad \hyperref[sec:next_gen_systems]{\textbf{Section VII: Integration with Emerging Technologies}} \\[2pt]
\quad \hyperref[sec:OTFS_Challenges]{\textbf{Section VIII: OTFS Channel Estimation Challenges}} \\[2pt]
\quad\quad $\triangleright$ \hyperref[subsec:leakage_suppression]{Leakage Suppression} \\[1pt]
\quad\quad $\triangleright$ \hyperref[subsec:inter_doppler_interference]{Inter-Doppler Interference} \\[1pt]
\quad\quad $\triangleright$ \hyperref[subsec:impulsive_noise]{Impulsive Noise} \\[1pt]
\quad\quad $\triangleright$ \hyperref[subsec:signalling_overhead]{Signalling Overhead} \\[1pt]
\quad\quad $\triangleright$ \hyperref[subsec:guard_space]{Guard Space Requirements} \\[1pt]
\quad\quad $\triangleright$ \hyperref[subsec:PAPR]{Peak-to-Average Power Ratio} \\[1pt]
\quad\quad $\triangleright$ \hyperref[subsec:beam_squint]{Beam Squint Effects} \\[1pt]
\quad\quad $\triangleright$ \hyperref[subsec:hardware_impairments]{Hardware Impairments} \\
\hline
\end{tabular}
\end{table}
\section{Background of OTFS Modulation}
\label{sec:background}
A basic OTFS system model with a single transmitter (Tx) and a single receiver (Rx) describes the basic components and signal processing stages of OTFS modulation, including the transformation of information symbols from the DD domain to the TF domain at the Tx, their propagation through the wireless channel, and the corresponding inverse operations at the Rx for data recovery.
\subsection{Transmitter}
\label{subsec:transmitter}
In OTFS modulation, $MN$ information symbols are multiplexed in the DD domain, represented as $x[k,l]$ where $k = 0, \ldots, N-1$ denotes the Doppler index and $l = 0, \ldots, M-1$ represents the delay index. The system parameters $M$ and $N$ specify the number of delay and Doppler bins, respectively. The DD grid is defined as $\Gamma = \left( \frac{k}{NT},\frac{l}{M \Delta f} \right)$, where $\Delta f = \frac{1}{T}$ represents the subcarrier spacing, $M\Delta f$ corresponds to the system bandwidth, and $NT$ denotes the OTFS frame duration \cite{8835764}.

The transformation from the DD domain to TF domain is obtained by applying the inverse symplectic finite Fourier transform (ISFFT), converting DD symbols $x[k,l]$ into TF symbols $X[n,m]$. The TF domain representation is expressed as \cite{8424569}
\begin{equation}
X\left [ n,m \right ] = \frac{1}{\sqrt{MN}}\sum_{k=0}^{N-1}\sum_{l=0}^{M-1}x\left [ k,l \right ]e^{j2\pi\left ( \frac{nk}{N}-\frac{ml}{M} \right )}.
\label{eq:1}
\end{equation}

Subsequently, the TF domain symbols are converted to a continuous-time signal $s(t)$ using the Heisenberg transform
\begin{equation}
s\left( t \right) = \sum_{n=0}^{N-1}\sum_{m=0}^{M-1}X\left[n,m\right]g_{\text{tx}}\left(t-nT  \right)e^{j2\pi m\Delta f\left( t-nT \right)},
\label{eq:2}
\end{equation}
where $g_{\text{tx}}\left( \cdot  \right)$ denotes the transmitter pulse-shaping function with duration $T$.

For practical implementation, the continuous-time signal $s(t)$ is discretized by sampling at intervals of $T/M$, yielding the discrete-time representation:
\begin{equation}
\mathbf{s} = \left[ s\left( 0 \right), s\left( 1 \right), \ldots, s\left( MN-1 \right) \right]^T.
\label{eq:3}
\end{equation}
The DD domain symbols can be arranged in matrix form as $\mathbf{X} \in \mathbb{C}^{M \times N}$:

\begin{equation}
\mathbf{X} =
\begin{bmatrix}
x[0,0] & \cdots & x[N-1,0]\\
\vdots & \ddots & \vdots\\
x[0,M-1] & \cdots & x[N-1,M-1]
\end{bmatrix}.
\label{eq:XcompactB}
\end{equation}
The transmitted TF domain symbol matrix $\mathbf{S} \in \mathbb{C}^{M \times N}$ is obtained as
\begin{equation}
\mathbf{S} = \mathbf{G}_{\text{tx}}\mathbf{F}^{{\dagger}}_M\left( \mathbf{F}_M \mathbf{X} \mathbf{F}^{{\dagger}}_N\right)= \mathbf{G}_{\text{tx}}\mathbf{X} \mathbf{F}^{{\dagger}}_N,
\label{Eq:5}
\end{equation}
where $\mathbf{G}_{\text{tx}}
=\operatorname{diag}\!\left(
g_{\text{tx}}(0),\allowbreak\, 
g_{\text{tx}}(T/M),\allowbreak\, 
\ldots,\allowbreak\,
g_{\text{tx}}((M-1)T/M)
\right)\allowbreak\,\mathrel{\in}\,\mathbb{C}^{M\times M}$ denotes the pulse-shaping matrix. For rectangular pulse shaping, $g_{\text{tx}}(t)=1$ for $0\le t\le T$ and $0$ otherwise, which yields $\mathbf{G}_{\text{tx}}=\mathbf{I}_M \in \mathbb{C}^{M\times M}$. The matrices $\mathbf{F}_N \in \mathbb{C}^{N\times N}$ and $\mathbf{F}_M \in \mathbb{C}^{M\times M}$ denote the normalized DFT matrices of size $N$ and $M$, respectively. The operator $(\cdot)^\dagger$ denotes the Hermitian transpose.

The relationship between the discrete-time signal vector $\mathbf{s}$ and the transmitted symbol matrix $\mathbf{S}$ is given by
\begin{equation}
\mathbf{s} = \text{vec}\left( \mathbf{S} \right) = \left( \mathbf{F}^{\dagger}_N \otimes \mathbf{G}_{\text{tx}}  \right)\mathbf{x},
\label{eq:6}
\end{equation}
where $\mathbf{x} = \text{vec}(\mathbf{X}) \in \mathbb{C}^{MN \times 1}$ is the vector form of the DD domain symbols, and $\otimes$ denotes the Kronecker product.
\subsection{Channel}
\label{subsec:channel}
The transmitted signal $s(t)$ propagates through a time-varying wireless channel characterized by the DD channel response.
For a multipath channel with $P$ propagation paths, the channel response is given by
\begin{equation}
h\left( \tau, \nu \right) = \sum_{p=1}^{P}h_p\delta\left( \tau-\tau_p \right)\delta\left( \nu-\nu_p \right),
\end{equation}
where $h_p$ denotes the complex channel gain of the $p$-th path, while $\tau_p$ and $\nu_p$ represent the corresponding delay and Doppler shift. The path parameters are discretized on the DD grid as $\Gamma_p = \left( \tau_p,\nu_p \right) = \left( l_p/{M\Delta f},{k_p}/{NT} \right)$. The time-domain channel matrix $\mathbf{H} \in \mathbb{C}^{MN \times MN}$ is defined as
\begin{equation}
\mathbf{H} \triangleq \sum_{p=1}^{P} h_p\,\mathbf{\Pi}^{\,l_p}\mathbf{\Delta}^{\,k_p+\kappa_p},
\label{eq:11}
\end{equation}
where $\mathbf{\Pi}$ is the permutation matrix modeling delay shifts and $\mathbf{\Delta}$ is the diagonal matrix modeling Doppler shifts, while $l_p$, $k_p$, and $\kappa_p$ denote the delay tap index, the integer Doppler index, and the fractional Doppler shift of the $p$-th path, respectively.
The permutation matrix $\mathbf{\Pi}$ is structured as follows:
\begin{equation}
   \mathbf{\Pi} =
   \begin{bmatrix}
       0 & \cdots & 0 & 1 \\
       1 & \ddots & 0 & 0 \\
       \vdots & \ddots & \ddots & 0 \\
       0 & \cdots & 1 & 0
   \end{bmatrix}_{MN \times MN}\hspace{-4em}.
\end{equation}
The diagonal matrix $\mathbf{\Delta}$ contains elements $\alpha = e^{\frac{j2\pi}{MN}}$ and is expressed as $\mathbf{\Delta} = \mathrm{diag}\left\{\alpha^0, \alpha^1, \ldots , \alpha^{MN-1} \right\}$.
\subsection{Receiver}
\label{subsec:receiver}
\begin{figure*}
    \centering    \includegraphics[width=.75\linewidth]{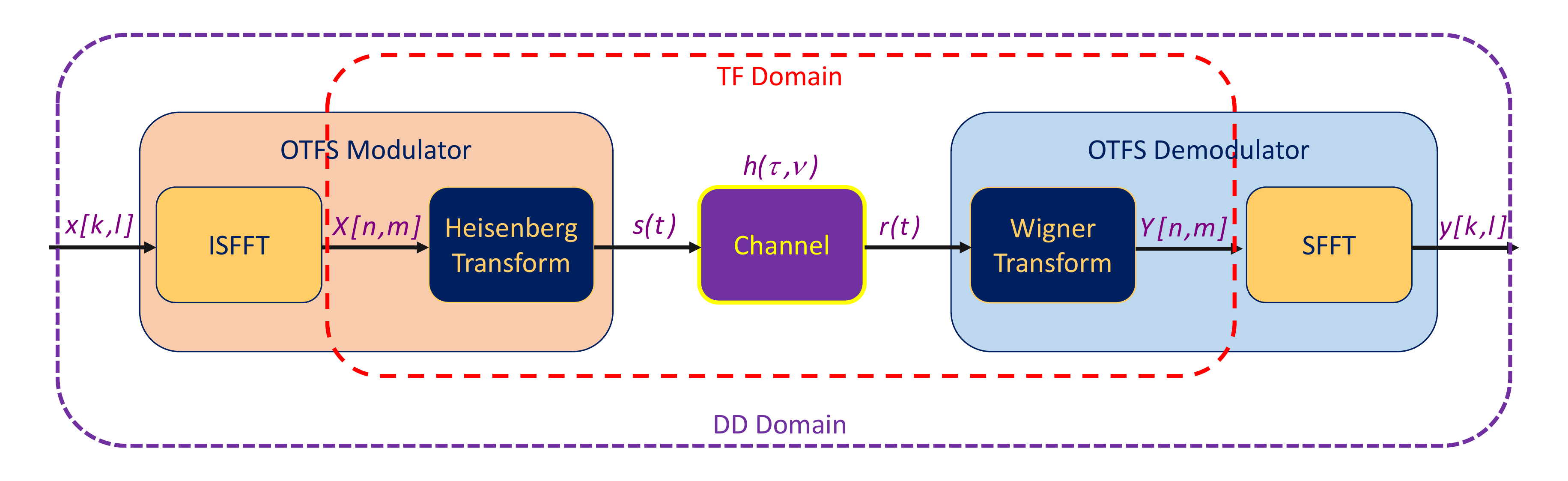}
    \caption{OTFS system model.}
    \label{fig:system}
\end{figure*}

The complete OTFS system model including the Tx, channel, and Rx is illustrated in Fig.~\ref{fig:system}. At the Rx, the received signal $r(t)$ is expressed as
\begin{equation}
r\left ( t \right ) = \int_{\tau} \int_{\nu} h\left ( \tau, \nu \right )s\left ( t-\tau \right )e^{j2\pi \nu \left ( t-\tau \right )}d \tau d \nu +  w\left ( t \right ),
\label{eq:8}
\end{equation}
where $w(t)$ represents the additive white Gaussian noise (AWGN). 

The receiver performs matched filtering on $r(t)$ using the Wigner transform, implemented via the cross-ambiguity function. The cross-ambiguity function between the receiver pulse-shaping function $g_{\text{rx}}(t)$ and the received signal $r(t)$ is defined as
\begin{equation}
\mathcal{A}_{g_{\text{rx}},r}(\tau,\nu) = \int g^*_{\text{rx}}(t-\tau)r(t)e^{-j2\pi\nu(t-\tau)}\mathrm{d}t.
\label{eq:cross_ambiguity}
\end{equation}

This continuous cross-ambiguity function is sampled at $\tau=nT$ and $\nu=m\Delta f$ to obtain the discrete TF domain symbols:
\begin{equation}
Y\left[n,m\right] = \mathcal{A}_{g_{\text{rx}},r}(nT, m\Delta f).
\label{eq:wigner_sampled}
\end{equation}

The continuous signal $r(t)$ is sampled at frequency $f_s = M/T$ to obtain the discrete signal as
\begin{equation*}
\mathbf{r} = \left[ r\left( 0 \right), r\left( 1 \right), \ldots, r\left( MN-1 \right) \right]^T =\left[ r\left( n \right) \right]_{n=0}^{MN-1}.
\end{equation*}
The discrete received signal can be expressed in terms of the channel parameters as
\begin{equation}
r\left( n \right) = \sum_{p=1}^{P}h_pe^{j2\pi\frac{k_p\left( n-l_p \right)}{MN}}s\left( \left[ n-l_p \right]_{MN} \right)+w\left( n \right),
\label{eq:9}
\end{equation}
where $\left[\cdot \right]_n$ denotes the modulo-$n$ operation.
In vector form, the received signal is represented as
\begin{equation}
   \mathbf{r} = \mathbf{H} \mathbf{s} + \mathbf{w},
\end{equation}
where $\mathbf{r},\mathbf{s}$ and $\mathbf{w}\in\mathbb{C}^{MN\times 1}$ denote the complex received vector, the transmitted vector and the noise vector, respectively. 

At the receiver, the transmitter operations are reversed to recover the transmitted symbols. First, the received discrete-time samples $\mathbf{r}$ are reshaped into the TF domain matrix $\mathbf{R} = \text{vec}^{-1}(\mathbf{r})$, where the vectorized samples are reorganized into an $M \times N$ matrix structure. Subsequently, the DD domain symbol matrix is obtained through $\mathbf{Y} = \mathbf{F}_M^\dagger (\mathbf{F}_M \mathbf{G}_{\text{rx}} \mathbf{R})\mathbf{F}_N$ \cite{8516353}. This demodulation process consists of an $M$-point FFT operation followed by the SFFT. The receiver pulse-shaping matrix $\mathbf{G}_{\text{rx}} \in \mathbb{C}^{M \times M}$ incorporates the matched filter based on the pulse-shaping waveform $g_{\text{rx}}(t)$, and is defined as $\mathbf{G}_{\text{rx}} = \text{diag}\{g_{\text{rx}}(0), g_{\text{rx}}(T/M), \ldots, g_{\text{rx}}((M-1)T/M)\}$. The complete input-output relation in the DD domain can be expressed in vector form as
\begin{align}
   \mathbf{y} &= (\mathbf{F}_N \otimes \mathbf{G}_{\text{rx}}) \mathbf{r} \nonumber \\
   &= (\mathbf{F}_N \otimes \mathbf{G}_{\text{rx}}) \mathbf{H} (\mathbf{F}_N^\dagger \otimes \mathbf{G}_{\text{tx}}) \mathbf{x}  
   + (\mathbf{F}_N \otimes \mathbf{G}_{\text{rx}}) \mathbf{w} \nonumber \\
   &= \mathbf{H}_{\text{eff}} \mathbf{x} + \tilde{\mathbf{w}},
\label{eq:inputoutput}
\end{align}
where $\mathbf{H}_{\text{eff}} = (\mathbf{F}_N \otimes \mathbf{G}_{\text{rx}}) \mathbf{H} (\mathbf{F}_N^\dagger \otimes \mathbf{G}_{\text{tx}})$ denotes the effective channel matrix, $\tilde{\mathbf{w}} = (\mathbf{F}_N \otimes \mathbf{G}_{\text{rx}}) \mathbf{w}$ represents the transformed noise vector.

\section{OTFS Channel Estimation Techniques}
\label{sec:CH_ES}
Reliable communication in modern wireless systems is critical to accurate CE, which is even more important in the context of OTFS modulation. As wireless communication systems expand to meet increasingly difficult propagation settings defined by high mobility, severe multipath effects, and quickly changing channel conditions, CE has emerged as an important predictor of overall system performance. With the goal of addressing issues with traditional OFDM for situations involving severe Doppler shifts, OTFS modulation converts the TF domain into the DD domain. In this transformed domain, the wireless channel appears sparse and nearly static, prompting a significant change in the approach to CE. The growing focus on OTFS-CE underlines its critical role in the growth of wireless communication technologies, as well as its potential to enable future advances in the field. This growing scientific interest is not solely driven by theoretical curiosity; it also clearly demonstrates the need to address real-world challenges in next-generation applications such as vehicle-to-everything (V2X) communication, high-speed train systems, UAV networks, and satellite-based communication. In these cases, relative movement speeds of hundreds of kilometers per hour need the adoption of advanced CE approaches.

The key advantage of OTFS modulation is its ability to convert a doubly-dispersive wireless channel into a two-dimensional (2D) representation in the DD domain, where channel parameters remain nearly constant throughout the transmission frame \cite{OTFSURVEY1}. This transformation offers several key advantages for CE. First, the sparse representation of the channel in the DD domain significantly reduces the number of parameters that need to be estimated, as practical wireless channels typically consist of only a few dominant propagation paths \cite{OTFSURVEY1}. Second, since the DD channel is quasi-static during an OTFS frame, a single CE suffices for the whole frame, drastically minimizing pilot overhead compared to OFDM, which requires frequent CE updates \cite{Raviteja_2019}. Furthermore, the OTFS framework provides a foundation for advanced estimation algorithms to take advantage of, leading to improved performance, due to the inherent relationship between delay and Doppler parameters \cite{Khan_2024}, \cite{hadani2018otfsnewgenerationmodulation}.

A rigorous evaluation of CE algorithms for OTFS systems under a variety of propagation environments and system configurations necessitates the development of theoretical lower bounds that define the minimum achievable estimation error. In this regard, \cite{9524512} derives the Cramér-Rao Lower Bound (CRLB) specifically for OTFS-CE, offering valuable theoretical benchmarks for both single-input single-output (SISO) and MIMO-OTFS scenarios. This enables comparative analysis and fair assessment of practical estimation algorithms across a wide spectrum of conditions.

A number of different methodological approaches have emerged from the important studies on OTFS-CE; all aim to solve various challenges in high-mobility wireless communications. According to the literature, research efforts are focused in a number of key areas. It has been found that BL techniques have significantly improved the estimation of fractional parameters and the recovery of sparse signals \cite{10292645,9110823,9686700,10711236,9771955,9483694,Wang_2023,Zhang_2024,10506450,9738478,11072479,10038838,9184852,9880804,10475894,10677422,10436557,10704024,10758799}. While MP-based techniques have demonstrated effectiveness in exploiting channel sparsity with a lower level of computational complexity \cite{8727425,10213984,9128497,9200896,10001420,9685856,9590508,9891774,10201832}. Methods based on MEP have made iterative estimation possible with nearly optimal performance \cite{9456029, Zhang_2023, 10310071, 9864300, 10870378}, and methods based on DL have brought data-driven solutions for challenging channel environments \cite{9896637,10138432,10439989,10400942,10540188, 10654761, 10856399, 10747184, 10916513, 10994238,11010104, 11071963}. Recent approaches target particular challenges, such as low-complexity estimation, synchronization, and hardware impairments \cite{10122554,10666708,10557727,Gui_2024,10582886,Guo_2024,10612829,10073950,Muppaneni_2023,9400868,10385066,Priya_2024,10742117,10816508,10938192,10993437}. Classification of CE methodologies is simplified by their operational domains and pilot structures. DD domain techniques are the predominant approach, using separate pilot \cite{8647394,8503182}, embedded pilot (EP) \cite{8671740,9303350,9794710,10700683,Zhao_2020}, and superimposed pilot (SP) schemes \cite{9539066,Liu_2023,10640141,10937496,10752433} to exploit sparse channel representation. TF domain methods offer alternative approaches like sequential delay, Doppler, and gain estimation \cite{10151793}. The integration of joint CE and SD has yielded promising results for improving overall system performance \cite{10138088,10147252,10887291,9785832,10475591,10476726,Yang_2024,9864300,9928043,RISOTFS14,10683792,11072043}.
OTFS-CE is being adapted for next-generation wireless systems, including ISAC \cite{10409528,Muppaneni_2023,10714398,10630836,11053775,11058951,11068135,fd_otfs}, RISs \cite{9864300,RISOTFS21,10976498,11010104}, massive MIMO \cite{9827947,10272673,10472131,10915571,10690180}, and UAV communications \cite{10815946}. 
Despite considerable progress, a number of major integration limitations remain. These include leakage \cite{9737331}, IDI mitigation \cite{He2023ATC, 11023080}, robustness against impulsive noise \cite{10209369}, signaling overhead reduction \cite{10223427, 10356117, 10370744, 10587305}, guard space minimization \cite{10264119, 10637960}, PAPR reduction \cite{10550432, 10614842, 11030215}, compensation for beam squint effects \cite{10970022}, and mitigation of hardware impairments \cite{11014223}. Overcoming these challenges can only be achieved through continuous innovation in OTFS-CE systems, which can address emerging concerns and fully exploit the potential of high-mobility wireless communications.

In OTFS modulation, information symbols are multiplexed in the DD domain, and the channel response exhibits sparse and long-term time-invariant properties in this domain, which greatly simplifies CE both conceptually and practically. On the receiver side, the accurate estimation of the amplitude and phase values corresponding to these delay and Doppler parameters is critical for successful symbol detection. Although these conceptual and practical advantages are available in the DD domain, practical OTFS systems encounter significant challenges when subjected to channel conditions that change rapidly during time. The inherent sparsity of the DD domain is disrupted by fractional Doppler effects, making standard estimation procedures inefficient. Several studies have been conducted to solve these challenges \cite{9440710}.

To exploit the sparse structure of the DD domain while mitigating these challenges, CE techniques in the literature are generally grouped into three main categories: the separate pilot approach, which uses separate pilot frames consisting solely of pilot symbols; the EP approach, in which pilot and data symbols are embedded in the same frame; and the SP approach, in which pilot symbols are superimposed on data symbols. The common goal of these methods is to reduce both the pilot load and computational complexity while maintaining high CE accuracy. These three approaches are described in detail in the following subsections.
\subsection{Separate Pilot Technique}
\label{subsec:separate_pilot}
\noindent
In this technique, the pilot and data frames are handled separately and the channel matrix estimated using the pilot frames is used to obtain the data frames. Separate pilot techniques can be further divided into three basic approaches.

\subsubsection{Impulse-based CE}
\label{subsubsec:impulse_based}
In this approach, impulses are transmitted as pilots \cite{8647394}. A single pilot is placed at a known DD-grid location \((k_{\text p},l_{\text p})\):
\begin{equation}
  x_\text{p}\left[ k,l \right] = 
    \begin{cases}
      1, &  \left( k,l \right) = \left( k_\text{p}, l_\text{p} \right),\\
      0, & \text{otherwise}.
    \end{cases}       
\end{equation} 
The effective windowed channel $h_w(\tau',\nu')$ is obtained by the circular convolution as follows:
\begin{equation}
h_w(\tau',\nu')=\int_{\tau}\!\int_{\nu} h(\tau,\nu)\,w_f(\tau'-\tau,\nu'-\nu)\,d\tau\,d\nu,
\end{equation}
where $\tau' = (k-n)/(NT)$ and $\nu' = (l-m)/(M\Delta f)$. The windowing function $w_f(\tau,\nu)$ is given by
\begin{equation}
w_f(\tau,\nu)=\sum_{n=0}^{N-1}\sum_{m=0}^{M-1}
W_{\rm tx}[n,m]\,W_{\rm rx}[n,m]\,
e^{-j2\pi(\nu nT-\tau m\Delta f)},
\label{eq:window_function}
\end{equation}
where $W_{\rm tx}[n,m]$ and $W_{\rm rx}[n,m]$ are the transmit and receive windowing functions on the TF grid, respectively. The received pilot frame in the DD domain can be obtained using 2D periodic convolution with \(h_w(\cdot,\cdot)\) as follows:
\begin{align}
y_{\text p}[k,l]
&=\frac{1}{MN}\sum_{k'=0}^{N-1}\sum_{l'=0}^{M-1}
x_{\text p}[k',l']\,
h_w\!\left(\frac{k-k'}{NT},\,\frac{l-l'}{M\Delta f}\right) \notag\\
&\quad + w[k,l].
\end{align}
If we transmit a pilot signal defined as \(x_{\text p}[k',l']=1\) for \((k',l')=(k_{\text p},l_{\text p})\) and \(0\) otherwise, the received pilot signal is given by
\begin{equation}
y_{\text p}[k,l]
=\frac{1}{MN}\,
h_w\!\left(\frac{k-k_{\text p}}{NT},\,\frac{l-l_{\text p}}{M\Delta f}\right)
+ w[k,l].
\label{eq:impulse_received}
\end{equation}
Since the pilot positions \((k_{\text p},l_{\text p})\) are known a priori at the receiver side, the windowed channel response \(\tfrac{1}{MN}\,h_w\!\left(\tfrac{k}{NT},\,\tfrac{l}{M\Delta f}\right)\) can be directly estimated from \eqref{eq:impulse_received}.
\subsubsection{Pseudo-Noise Based CE}
\label{subsubsec:pseudo_noise}
This method uses a pseudo-noise (PN) sequence as a pilot \cite{8503182}.
Like the impulse-based method, the PN approach also uses a dedicated pilot frame, i.e., a transmit pilot $x_\text{p}\left[ k,l \right]$ and the corresponding received pilot $y_\text{p}\left[ k,l \right]$. The key difference is the pilot structure and the estimator. Similar to the previously defined pilot symbol $x_\text{p}\left[ k,l \right]$, the PN pilot sequence is denoted by $x_\text{pn}\left[ k,l \right]$. It is a sequence of length $N_p$ transmitted during a pilot interval, and its corresponding received pilot is $y_\text{pn}\left[ k,l \right]$. Let \(\mathbb{Z}_{N_{\text{pn}}}=\{0,1,\ldots,N_{\text{pn}}-1\}\) denote the finite index set, and let \(\mathcal{H}=\{\,f:\mathbb{Z}_{N_{\text{pn}}}\!\to\!\mathbb{C}\,\}\) be the vector space of complex-valued functions on \(\mathbb{Z}_{N_{\text{pn}}}\).
Let \(s_{\text{pn}}[n]\in\mathcal{H}\), with \(M\ge N_{\text{pn}}\) and sampling period \(T_s=1/W\). The transmitted analog signal is then given by
\begin{equation}
s_A(t)=\sum_{n=0}^{M-1} s_{\text{pn}}\!\big[n \bmod N_{\text{pn}}\big]\;\mathrm{sinc}(W t - n).
\end{equation}
The coupling between the transmitted pilot waveform \(s_A(t)\) and the received waveform \(r_A(t)\) follows the continuous-time input–output relation in ~\eqref{eq:8}. Let \(T_{\text{spread}}=\max\{\tau_p\}\) denote the time spread of the channel and define \(K=\left\lceil T_\text{spread} W\right\rceil\,\). With \(M=N_{\text{pn}}+K\) and sampling period \(T_s=1/W\), the pilot-interval discrete observation is obtained by \(r_{\text{pn}}[n]=r_A\!\big((K+n)/W\big)\) for \(n=0,1,\ldots,N_{\text{pn}}-1\).

When the delays and Doppler shifts satisfy \(\tau_p\in \tfrac{1}{W}\mathbb{Z}_{+}\) and \(\nu_p\in \tfrac{W}{N_{\text{pn}}}\mathbb{Z}\), the sampled pilot observation obeys the discrete cyclic DD model with integer shifts
\begin{equation}
    r_{\text{pn}}[n]=\sum_{p=1}^{P} \alpha_p\, e^{j\frac{2\pi \omega_p}{N_{\text{pn}}}n}\, s_{\text{pn}}[\,n-\delta_p\,]+w[n].
    \label{eq:20}
\end{equation}
Here the discrete indices are \(\delta_p=\tau_p W\) and \(\omega_p=\tfrac{N_{\text{pn}}\nu_p}{W}\), corresponding to delay and Doppler, respectively; the effective complex coefficient collects the path gain and the sampling-offset phase, \(\alpha_p=h'_p e^{j2\pi \nu_p K/W}\) with \(h'_p=h_p e^{-j2\pi \nu_p \tau_p}\), where \(\alpha_p\in\mathbb{C}\), \(\delta_p,\omega_p\in\mathbb{Z}_{N_{\text{pn}}}\), and \(w[n]\in\mathcal{H}\).

Define a simpler variant of the expression in \eqref{eq:20} as follows:
\begin{equation}
r_{\text{pn}}[n]=e^{j\frac{2\pi \omega_0}{N_{\text{pn}}}n}\,s_{\text{pn}}[n-\delta_0]+w[n].
\end{equation}
Subsequently, the matched filter output is given by
\begin{equation}
\mathcal{M}[\delta,\omega]\;\triangleq\;\frac{1}{N_{\text{pn}}}\sum_{n=0}^{N_{\text{pn}}-1}
r_{\text{pn}}[n]\;s_{\text{pn}}^{*}[n-\delta]\;e^{-j\frac{2\pi \omega}{N_{\text{pn}}}n},
\label{eq:24}
\end{equation}
where $\delta,\omega\in\mathbb{Z}_{N_{\text{pn}}}$. If $s_{\text{pn}}$ is a unit-norm PN sequence and $N_{\text{pn}}\to\infty$ \cite{6563167}, we have
\begin{equation}
\mathcal{M}[\delta,\omega]=
\begin{cases}
1+\varepsilon'_{N_{\text{pn}}}, & (\delta,\omega)=(\delta_0,\omega_0)\\[2pt]
\varepsilon_{N_{\text{pn}}}, & (\delta,\omega)\neq(\delta_0,\omega_0),
\end{cases}
\label{eq:25}
\end{equation}
with error terms bounded by $\bigl|\varepsilon'_{N_{\text{pn}}}\bigr|\le N_{\text{pn}}^{-1/2}$ and $\bigl|\varepsilon_{N_{\text{pn}}}\bigr|\le (C+1)\,N_{\text{pn}}^{-1/2}$, where $C>0$ is a constant. 

Hence, the delay and Doppler estimates are obtained by finding the entries where $|\mathcal{M}[\delta,\omega]|$ is maximum:
\begin{equation}
(\hat{\delta}_p,\hat{\omega}_p)\;=\;\arg\max_{\delta,\omega}\,\big|\mathcal{M}[\delta,\omega]\big|,
\label{eq:26}
\end{equation}
and the path coefficient is estimated as $\hat{\alpha}_p=\mathcal{M}[\hat{\delta}_p,\hat{\omega}_p]$. For multiple paths ($P>1$), the $P$ largest peaks of $|\mathcal{M}[\delta,\omega]|$ are selected to estimate $\{(\hat{\delta}_p,\hat{\omega}_p)\}_{p=1}^P$.
\subsection{Embedded Pilot Technique}
\label{subsec:embedded_pilot}
In this technique, an entire frame is not allocated for pilot transmission. The pilot symbols are transmitted in the same frame as the data symbols and guard symbols are inserted to prevent interference between the pilot and data symbols \cite{naikoti2021signal}. In absence of guard symbols, other approaches are required to mitigate inter symbol interference (ISI).
\subsubsection{Embedded Pilot Based CE}
\label{subsubsec:embedded_estimation}
In this method, we initially determine an arbitrary grid position $[k_p, l_p]$ such that $0 \leq k_p \leq N - 1$ and $0 \leq l_p \leq M - 1$. Let $l_\tau$ and $k_\nu$ denote the maximum delay and Doppler shift, respectively. For ease of representation, we consider $0 \leq l_p - l_\tau \leq l_p \leq l_p + l_\tau \leq M - 1$ and $0 \leq k_p - 2k_\nu \leq k_p \leq k_p + 2k_\nu \leq N - 1$ \cite{Raviteja_2019}. In this case, the pilot, guard, and data symbols can be expressed as follows:
\begin{equation}
x[k, l] = 
\begin{cases}
    x_p[k, l], & \text{if } k = k_p, \; l = l_p, \\
    0, & \text{if } k_p - 2k_\nu \leq k \leq k_p + 2k_\nu, \\
       & \quad l_p - l_\tau \leq l \leq l_p + l_\tau, \\
    x_d[k, l], & \text{otherwise.}
\end{cases}
\end{equation}
\begin{figure*}[t!]
    \centering
    \includegraphics[width=0.75\textwidth]{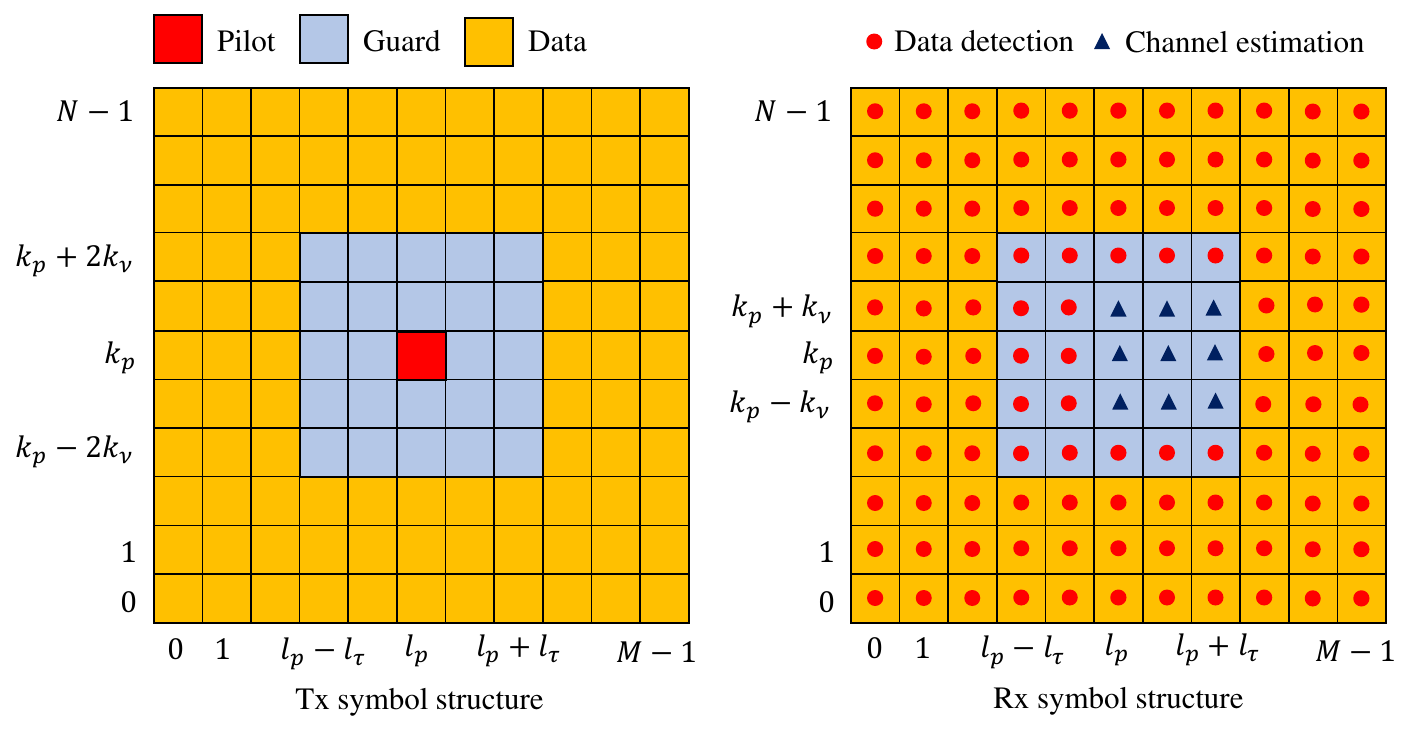}
    \caption{Tx and Rx symbol structures of the EP-based CE.}
    \label{fig:embedded CE}
\end{figure*}
The location of the symbol $y[k, l]$ used for CE in the frame is specified by the constraints $k_p - k_\nu \leq k \leq k_p + k_\nu$, and $l_p \leq l \leq l_p + l_\tau$. The remaining $y[k, l]$ symbols are used for data detection. For the integer Doppler case, the symbols used for CE and data detection are illustrated on the DD grid structure in Fig. \ref{fig:embedded CE}. A binary indicator \(b[k',l']\in\{0,1\}\) is used: \(b[k',l']=1\) indicates a path at Doppler tap \(k'\) and delay tap \(l'\), whereas \(b[k',l']=0\) indicates no path.
For \( k \in [0, N-1] \) and \( l \in [l_p, l_p + l_\tau] \), the estimated symbol \( y[k,l] \) is obtained as follows:
\begin{equation}
y[k, l] = b[l - l_p] h[[k - k_p]_N, l - l_p] x_p + v[k, l].
\end{equation}

Recent OTFS-CE research has increasingly converged on EP-based designs, owing to their ability to estimate the channel with minimal overhead while coexisting with data symbols on the DD grid. Study \cite{8671740} presents a comprehensive early work in OTFS-CE that forms the foundation for subsequent research in this field, proposing EP-CE schemes for OTFS modulation that arrange pilot, guard, and data symbols in DD plane to avoid interference, using threshold-based detection for both integer and fractional Doppler cases with minimum overhead and extensions to MIMO systems. In a related EP approach, the study \cite{9303350} suggests a DD domain EP-based CE method for mobile OTFS systems that deals with residual frame timing offset, carrier frequency offset, and fractional multiple Doppler effects. The approach demonstrates that time domain equivalent channel matrix structure is invariant to Doppler values and achieves minimum mean square error (MMSE) equalization performance approaching ideal knowledge-based systems in high-mobility scenarios. For code-domain non-orthogonal multiple access (NOMA) applications, \cite{9794710} develops an EP-aided CE technique for OTFS-sparse code multiple access (SCMA) systems using convolutional sparse coding (CSC) framework for uplink multi-user communications. Recent advances in EP techniques include \cite{10700683}, which introduces a pulsone-based framework that exploits the self-dual properties of pilot impulses between time and frequency domains, enabling fine-resolution estimation of fractional delay and Doppler parameters while maintaining low computational complexity.

\subsubsection{Sparse Bayesian Learning Based CE}
\label{subsubsec:sparse_bayesian}
\begin{figure*}[t!]
    \centering
    \includegraphics[width=0.75\textwidth]{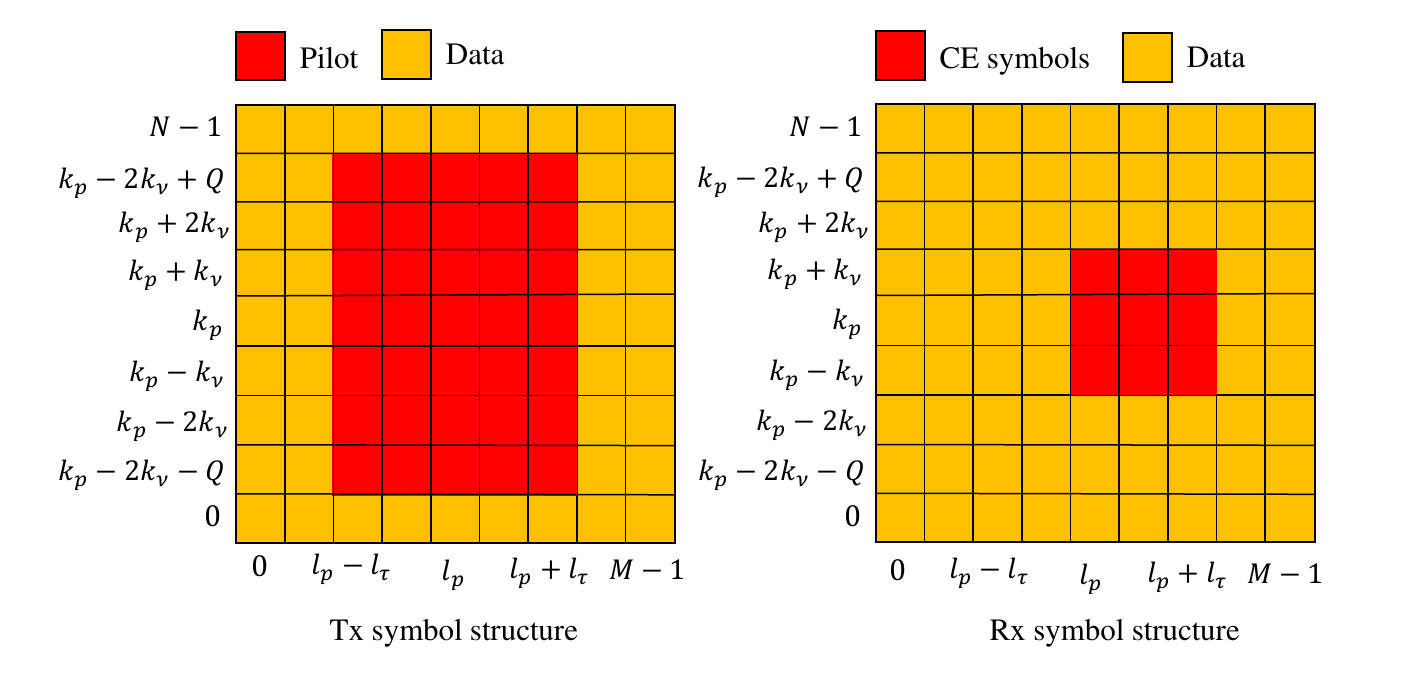}
    \caption{Tx and Rx symbol structures of the SBL-based CE.}
    \label{fig:SBL CE}
\end{figure*}

In this approach, the sparsity of the channel in the DD domain is exploited, and there is no need for guard symbols. Additionally, the power of pilot and data symbols is consistent, which reduces pilot overhead and improves pilot signal-to-noise-ratio (SNR) \cite{Zhao_2020}. In pilot design, a \( Q \) value is initially selected to reflect the effect of fractional Doppler. It has been observed that a \( Q \) value of approximately 5 yields successful results \cite{Raviteja_2019}. In this case, pilot and data symbols can be constructed as follows:
\begin{equation}
x[k, l] = 
\begin{cases} 
x_p[k, l] & k_p - 2 k_{\nu} - Q \leq k \leq k_p + 2 k_{\nu} + Q, \\ 
           & l_p - l_{\tau} \leq l \leq l_p + l_\tau, \\
x_d[k, l] & \text{otherwise}.
\end{cases}
\end{equation}
Setting $k_p = N/2$ and $l_p = M/2$ allows for achieving a balance between system performance and pilot overhead, depending on $l_\tau$ \cite{Zhao_2020}. The received symbols $y[k, l]$ for CE are selected such that $k \in [k_p - k_\nu, k_p + k_\nu]$ and $l \in [l_p, l_p + l_\tau]$. The Tx and Rx symbol structures related to SBL are illustrated in Fig. \ref{fig:SBL CE}.

\subsection{Superimposed Pilot Based Estimation Technique}
\label{subsec:superimposed_pilot}
This method involves the superimposition of pilot symbols onto data symbols within the OTFS frame \cite{9456894}. The pilot and data symbols can be defined as follows:
\begin{equation}
x[k, l] = \begin{cases}
    x_p[k, l] + x_d[k, l], & \text{if } k = k_p, \, l = l_p, \\
    x_d[k, l], & \text{otherwise.}
\end{cases}
\end{equation}
The pilot symbol $x_p[k, l]$ is superimposed on the data symbol $x[k_p, l_p]$, resulting in inevitable interference $I_{k,l}$:
\begin{equation}
y[k, l] = x_p[k, l] h \left[ (k - k_p)_N, (l - l_p)_M \right] + I_{k,l} + w[k, l], 
\end{equation}
where $I_{k,l}$ denotes the interference introduced by the data symbols, given by
\begin{equation}
I_{k,l} = \sum_{k' = k - k_\nu}^{k + k_\nu} \sum_{l' = l - l_\tau}^{l} x[k', l'] h \left[ (k - k')_N, (l - l')_M \right].
\end{equation}

Recent progress has shown how SP-based CE could improve the efficiency of OTFS systems. In particular, an effective data-aided CE algorithm has been proposed in \cite{9456894}, where both pilot and data symbols are transmitted within the OTFS frame, requiring only a single pilot symbol. This approach utilises iterative CE, interference cancellation and a sum-product algorithm (SPA)-based data detection method to mitigate the inherent pilot-data interference in SP schemes. Simulations show that the suggested method achieves bit error rate (BER) comparable to typical separate pilot systems while significantly improving spectral efficiency due to the efficient use of pilot resources. Alternative SP-based estimation and detection frameworks \cite{9539066} have been introduced for OTFS, focusing on maximizing spectral efficiency while maintaining reliable detection. These approaches superimpose low-powered pilots on data symbols in the DD domain, and MEP is used to exploit channel sparsity. The SP-NI (non-iterative) and SP-I (iterative) designs, in particular, stand out for exceeding standard EP techniques' spectral efficiency with a BER performance equivalent to existing schemes. In another SP-based study \cite{Liu_2023}, scattered SP were developed where multiple low-power pilot symbols are strategically distributed and superimposed with data symbols across the DD domain, enabling significant PAPR reduction while maintaining high spectral efficiency and achieving superior CE performance through joint processing of multiple independent received signals. Moreover study in \cite{10640141}, SP-based OTFS-CE have introduced a sophisticated framework that uses multiple-frame SP structures alongside SBL to take advantage of 2D DD channel sparsity. The multiple-frame coupled prior SBL (M-CPSBL) approach and its computationally efficient counterpart, multiple-frame-based-CPSBL (M-CPSBL), are designed to estimate the sparse channel while also mitigating mutual pilot-data interference. These methods operate on SP-OTFS systems in which low-power pilot symbols are superimposed onto data across multiple frames, thereby enabling iterative CE and data detection. To address doubly fractional channel conditions, a segmentation-based SBL (S-SBL) scheme has been developed for OTFS systems \cite{10937496}. The approach introduces DD segmentation factors adaptively selected based on channel quality to construct an extended measurement matrix, employing a 2D compressed sensing (CS) method that processes delay and Doppler dimensions separately with weighted average fusion to eliminate interference between fractional parameters while achieving significant complexity reduction. In addition, recent research has introduced a superimposed full pilots (SFP)-aided iterative CE (SFPCE) framework for multi-user OTFS uplink \cite{10752433}. The SFPCE approach, through the integration of maximum mean square error (MMSE)-based estimation, MEP detection, and an effective interference cancellation algorithm, demonstrates markedly enhanced BER and MSE performance compared to traditional SP-based methods, especially with an increasing number of users, all while preserving manageable computational complexity. This underscores the increasing potential of sophisticated SP-based iterative CE methodologies for next-generation OTFS systems.

\section{Time-Frequency Domain Channel Estimation}
\label{sec:tf_domain}
OTFS has emerged as a promising solution for communication over doubly selective channels; however, accurate CE in the TF domain remains relatively underexplored. Existing research on TF domain OTFS channel estimation is limited, with the work of Sheng et al. being particularly notable for its proposed pilot schemes and methodological contributions \cite{10151793}. TF domain CE (TF-CE) in OTFS is typically implemented using OFDM-based system frameworks in the existing literature \cite{8516353}, \cite{8119540}.
The DD domain symbols are transformed to the TF domain as
\begin{equation}
   \mathbf{X}_{\text{TF}} = \mathbf{F}_M \mathbf{X}_{\text{DD}} \mathbf{F}_N^{\dagger},
   \label{Eq:31}
\end{equation}
\noindent
where $\mathbf{X}_{\text{DD}} \in \mathbb{C}^{M \times N}$ represents the information symbols arranged in the DD grid. Note that the notational differences between \eqref{Eq:31} and \eqref{Eq:5} are introduced to enhance clarity in the OFDM-based OTFS framework, while both equations serve equivalent purposes.
The transmitted signal is obtained by vectorizing the inverse Fourier transformed TF symbols
\begin{equation}
    \mathbf{x} = \text{vec}\left(\mathbf{F}_M^{\dagger} \mathbf{X}_{\text{TF}}\right) = \text{vec}\left(\mathbf{X}_{\text{DD}} \mathbf{F}_N^{\dagger}\right).
\end{equation}
At the receiver, after cyclic prefix (CP) removal and OFDM demodulation, the DD domain symbols are recovered through
\begin{equation}
    \mathbf{R}_{\text{DD}} = \mathbf{F}_M^{\dagger} \left(\mathbf{F}_M \mathbf{R}\right) \mathbf{F}_N = \mathbf{X}_{\text{DD}}.
\end{equation}

In the frequency domain representation of OTFS modulation, pilot symbols are strategically allocated at specific subcarrier positions to facilitate CE. These pilot locations are characterized by the index set ${\mathcal{P}} = {{P}_1, {P}_2, \ldots, {P}_L} \subseteq {1, 2, \ldots, M}$, where $M$ denotes the total number of subcarriers. This analysis considers the case of uniformly distributed pilots, where the first pilot is positioned at ${P}_1 = 1$ and subsequent pilots follow the pattern ${P}_\ell = 1 + (\ell - 1)\varpi$ for $\ell = 2, 3, \ldots, L$, with $\varpi$ representing the constant spacing between adjacent pilot subcarriers. To ensure mathematical tractability, the total number of subcarriers can be expressed as $M = L\varpi$, which guarantees that pilots uniformly divide the available spectrum.

The elements of this pilot matrix are defined such that $[\mathbf{\tilde{S}}]_{m,n} \overset{\Delta}{=}\tilde{s}_{m,n}$ when $m \in \mathcal{P}$ and $n \in {1, 2, \ldots, N}$, while all other entries remain zero. Here, $\tilde{s}_{m,n}$ denotes the pilot symbol transmitted at the $m$-th subcarrier during the $n$-th OTFS symbol period. The time-domain transmitted pilot signal is obtained as
\begin{equation}
    \mathbf{\tilde{x}} = \text{vec}\left(\mathbf{F}_M^{\dagger} \mathbf{\tilde{S}}\right).
\end{equation}

Let $\mathbf{\hat{X}}$ and $\mathbf{\hat{S}}$ denote the data matrices in the DD and TF domains, respectively. Under the assumption of unit power normalization, where $\mathbb{E}\{|[\hat{X}]_{m,n}|^2\} = 1$ holds for all indices $m$ and $n$, where $\mathbb{E}\{\cdot\}$ denotes the expectation operator, we have
\begin{equation}
     \mathbf{\hat{S}} = \mathbf{F}_M \mathbf{\hat{X}} \mathbf{F}_N^{\dagger}.
\end{equation}
The resulting time-domain data signal is obtained as follows:
\begin{equation}
    \mathbf{\hat{x}} = \text{vec}\left(\mathbf{\hat{X}}\mathbf{F}_N^{\dagger}\right) \in \mathbb{C}^{MN \times 1}.
\end{equation}

After propagation through the channel and subsequent removal of the CP, the received signal can be mathematically represented as
\begin{equation}
r(t) = \sum_{p=1}^{P} h_p \, x\left([t - \tau_p]_{NT}\right) e^{j2\pi\nu_p t} + n(t), \quad t \in [0, NT]
\label{eq:r}
\end{equation}
The received signal in \eqref{eq:r} can be written in vector form as
\begin{equation}
\mathbf{r} = \tilde{\mathbf{r}} + \hat{\mathbf{r}} + \mathbf{n} \in \mathbb{C}^{MN \times 1},
\end{equation}
where $\tilde{\mathbf{r}}$ denotes the contribution from pilot symbols, $\hat{\mathbf{r}}$ represents the data symbol component, and $\mathbf{n}$ is the complex noise vector.
The total pilot signal component is obtained by summing the individual pilot contributions as $\tilde{\mathbf{r}} = \sum_{\ell=1}^{L} \tilde{\mathbf{r}}_\ell$. Therefore the effect of the known pilot symbols is compensated for as follows:
\begin{equation}
\mathbf{y}_\ell = \frac{M}{\tilde{s}_\ell} \, \mathrm{diag} \left\{ e^{-j2\pi\frac{P_\ell'}{M} \cdot 0},\, \ldots,\, e^{-j2\pi\frac{P_\ell'}{M}(MN-1)} \right\} \, \hat{\mathbf{r}}_\ell,
\label{eq:39}
\end{equation}
where $\tilde{s}_\ell$ denotes the transmitted pilot symbol, and $P_\ell' = P_\ell - 1$ accounts for the use of zero-based indexing in the subcarrier numbering. \eqref{eq:39} can be equivalently reformulated as
\begin{equation}
\mathbf{y}_\ell = \mathbf{A}(\boldsymbol{\nu}) \, \mathrm{diag}\{\mathbf{h}\} \begin{bmatrix}
e^{-j2\pi\frac{P_\ell'\tau_1}{T}} \\
\vdots \\
e^{-j2\pi\frac{P_\ell'\tau_P}{T}}
\end{bmatrix},
\end{equation}
where $\mathbf{A}(\boldsymbol{\nu}) \triangleq\mathbf{A} \in \mathbb{C}^{MN \times P}$ is the Doppler matrix with elements $[\mathbf{A}(\boldsymbol{\nu})]_{i,p} = e^{j2\pi\nu_p T(i-1)/M}$ for $i \in \{1, 2, \ldots, MN\}$ and $p \in \{1, 2, \ldots, P\}$, and $\mathbf{h} \in \mathbb{C}^{P\times1}$ is the vector of complex channel gains.
Combining all $L$ pilot observations yields
\begin{equation}
\mathbf{Y} = [\mathbf{y}_1,\, \ldots,\, \mathbf{y}_L] = \mathbf{A} \, \mathrm{diag}\{\mathbf{h}\} \, \mathbf{B}^T(\tau) \in \mathbb{C}^{MN\times L},
\end{equation}
where $\mathbf{B}^T(\tau) \triangleq \mathbf{B}^T \in \mathbb{C}^{P \times L}$ is the delay matrix with entries $[\mathbf{B}^T]_{p,\ell} = e^{-j2\pi P_\ell'\tau_p/T}$.

Let $\mathbf{\Phi}_p = \left[
    \mathrm{diag}\{\mathbf{g}_{1,p}\} \mathbf{\Psi}_p,\, \ldots,\, \mathrm{diag}\{\mathbf{g}_{N,p}\} \mathbf{\Psi}_p
\right]$ be the matrix associated with the $p$-th propagation path, where $\mathbf{g}_{n,p} \in \{0,1\}^{MN \times 1}$ is an indicator vector with $[\mathbf{g}_{n,p}]_i = 1$ for $i \in \{[\lceil \tau_p/T_s \rceil + (n-1)M]_{MN} + 1, \ldots, [\lceil \tau_p/T_s \rceil + nM - 1]_{MN} + 1\}$ and $[\mathbf{g}_{n,p}]_i = 0$ otherwise, where $T_s$ denotes the sampling period. Accordingly, the received data vector can be expressed as
\begin{equation}
    \hat{\mathbf{r}} = \mathbf{\Phi} \, \mathrm{vec}(\hat{\mathbf{S}}),
\end{equation}
where the overall channel matrix is given by
\begin{equation}
    \mathbf{\Phi} \overset{\Delta}{=} \sum_{k=1}^{P} h_p \mathbf{\Phi}_p \in \mathbb{C}^{MN \times MN}.
\end{equation}
The relationship between the channel using pilot symbols and the channel operating on data symbols can be expressed as follows:
\begin{equation}
\mathbf{C}\mathbf{h} = \mathbf{\Phi}\mathrm{vec}(\hat{\mathbf{S}}),
\end{equation}
where $\mathbf{C}$ denotes the pilot-based CE matrix that incorporates both delay and Doppler effects acting upon pilot symbols. 

Since pilots occupy only $1/Q$ of the bandwidth due to their $Q$-spaced frequency allocation, a $Q$-fold down-sampling can be applied. After truncating $N_0$ edge samples and retaining $M_0 \ll MN$ samples, the down-sampling index set becomes $\mathcal{M} \triangleq \{N_0 + qQ\}_{q=0}^{M_0-1} \subset \{1,\ldots,MN\}$, significantly reducing computational complexity in subsequent processing.

The extracted pilot signals from all $L$ subcarriers are arranged into a combined matrix $\hat{\mathbf{Y}} \overset{\Delta}{=} \left[ \hat{\mathbf{y}}_1, \ldots, \hat{\mathbf{y}}_L \right] \in \mathbb{C}^{MN \times L}$, which can be approximated as $\hat{\mathbf{Y}} \approx \mathbf{A} \mathrm{diag}\{\mathbf{h}\} \mathbf{B}^T$. The approximation arises from imperfect isolation during frequency-domain windowing.
The pilot matrix is down-sampled according to the index set $\mathcal{M}$, yielding $\hat{\mathbf{Y}}_0 \overset{\Delta}{=} \left[ \hat{\mathbf{Y}} \right]_{\mathcal{M}} \in \mathbb{C}^{M_0 \times L}$. This down-sampled matrix can be expressed as $\hat{\mathbf{Y}}_0 \approx \mathbf{A}_0 \mathrm{diag}\{\mathbf{h}\} \mathbf{B}^T$, with $\mathbf{A}_0 = \left[ \mathbf{A} \right]_{\mathcal{M}}$ being the corresponding down-sampled Doppler matrix.

\subsection{Delay Estimation}
\label{subsec:delay_estimation}
The delay estimation is performed using a correlation matrix constructed from the down-sampled pilot observations. The delay correlation matrix is defined as \begin{equation} \boldsymbol{\Sigma}_{\tau} \overset{\Delta}{=} \hat{\mathbf{Y}}_0^T \hat{\mathbf{Y}}_0^* = \mathbf{B}(\boldsymbol{\tau}) \boldsymbol{\Sigma}_{\mathbf{Ah}} \mathbf{B}^\dagger(\boldsymbol{\tau}) + \boldsymbol{\Xi}, \end{equation} where $\boldsymbol{\Xi}$ captures the noise contributions, and the matrix $\boldsymbol{\Sigma}_{\mathbf{A}\mathbf{h}} \triangleq \mathrm{diag}\{\mathbf{h}\} \mathbf{A}_0^T \mathbf{A}_0^* \mathrm{diag}\{\mathbf{h}^*\}$ encapsulates coupled Doppler-gain effects. A key property of this formulation is that the signal subspace of the correlation matrix $\boldsymbol{\Sigma}_{\tau} \in \mathbb{C}^{L \times L}$ depends solely on the delay parameters $\tau_p$, provided the Doppler and gain terms maintain full rank. This decoupling facilitates the application of subspace-based methods such as root-MUSIC for accurate delay estimation \cite{10151793}.
\subsection{Doppler Estimation}
\label{subsec:doppler_estimation}
Following the delay estimation, the Doppler shifts are determined through phase analysis of the reconstructed channel matrix. With the estimated delays $\hat{\boldsymbol{\tau}}$, the channel coefficient matrix is obtained as
\begin{equation}
\mathbf{A}_{\mathbf{h}} \overset{\Delta}{=} \hat{\mathbf{Y}}_0 [\mathbf{B}^T(\hat{\boldsymbol{\tau}})]^{\ddagger},
\end{equation}
where $[\cdot]^{\ddagger}$ denotes the Moore-Penrose pseudoinverse. For the \(p\)-th path, the vector \( [\mathbf{A}_{\mathbf h}]_{p} \) is defined as
\begin{equation}
[\mathbf{A}_{\mathbf{h}}]_{p} = h_p \begin{bmatrix}
e^{j2\pi\frac{T\nu_p(N_0-1)}{M}} \\
e^{j2\pi\frac{T\nu_p(N_0-1+Q)}{M}} \\
\vdots \\
e^{j2\pi\frac{T\nu_p(N_0-1+(M_0-1)Q)}{M}}
\end{bmatrix}.
\end{equation}
The Doppler shift $\nu_p$ for each path is estimated by extracting the phase slope through line fitting. Specifically, the phase of the $m$-th element follows $\phi_m = 2\pi\frac{T\nu_p}{M}(N_0 - 1 + (m-1)Q) + \phi_0$, where $\phi_0$ represents the initial phase offset. This linear relationship enables robust Doppler estimation through least-squares fitting of the unwrapped phase values \cite{10151793}.

\subsection{Channel Gain Estimation}
\label{subsec:channel_gain_estimation}
Upon obtaining the delay and Doppler estimates, the channel gains are determined through a least-squares approach. The reconstructed channel matrix is formulated as
\begin{equation}
\hat{\mathbf{C}} \overset{\Delta}{=} \mathbf{C}(\hat{\boldsymbol{\tau}}, \hat{\boldsymbol{\nu}}),
\end{equation}
where $\hat{\boldsymbol{\tau}}$ and $\hat{\boldsymbol{\nu}}$ denote the estimated delay and Doppler vectors, respectively. The channel gain vector is subsequently estimated via
\begin{equation}
\hat{\mathbf{h}} = \hat{\mathbf{C}}^{\ddagger} \mathbf{r},
\end{equation}
where $\mathbf{r}$ represents the received signal vector and $\hat{\mathbf{C}}^{\ddagger}$ is the pseudoinverse of the reconstructed channel matrix. This least-squares solution minimizes the estimation error $\|\mathbf{r} - \hat{\mathbf{C}}\hat{\mathbf{h}}\|^2$, providing optimal channel gain estimates under the assumption that $MN \gg K$. The overdetermined nature of the system ensures robust gain estimation without requiring explicit MMSE formulation.

\section{Signal Processing Algorithms for Channel Estimation}
\label{sec:ce_algorithms}
Accurate CE is critical to the performance of OTFS systems, and many signal processing algorithms have been developed to address this issue. This section divides these approaches into five categories: BL methods, MP algorithms, MEP techniques, DL approaches, and recent techniques that provide alternative perspectives.
\subsection{Bayesian Learning}
\label{subsec:bayesian_learning}
BL provides a probabilistic framework for CE that combines prior knowledge of channel statistics with updates in response to newly acquired observations. It accounts for uncertainty and adapts to changing conditions over time by describing channel parameters as random variables. This strategy enhances mean squared error, robustness to model mismatches, and computing efficiency, especially when pilot symbols are constrained or channels change rapidly. For multi-user OTFS systems, \cite{10292645} introduces a variational BL (VBL) approach that exploits shared sparsity across users through an effective Gaussian mixture prior, developing structured variational distributions to address posterior distribution intractability and significantly outperforming conventional modified subspace pursuit (MSP) and SBL-based estimators. Study \cite{9110823} presents an optimized high-mobility downlink CE scheme for massive MIMO-OTFS networks that formulates a time-domain signal model and adopts expectation maximization (EM)-based VBL (EM-VBL) framework to recover uplink channel parameters, then exploits angle-delay-Doppler reciprocity to reconstruct downlink channels with significantly reduced complexity through fast Bayesian inference. In \cite{9686700}, an efficient CE approach for MIMO-OTFS systems is proposed. Compared to earlier methods, this method achieves better performance and noise resilience by combining iterative block reorganization (BR) with block sparse BL (BSBL), resulting in delay path clusters capable of handling multiple Doppler shifts. The adaptive pattern-coupled SBL algorithm, introduced by \cite{10711236}, improves estimation accuracy over BSBL methods by using adaptive hyperparameter strategies and hierarchical Gaussian prior models to characterize relationships between adjacent coefficients. In \cite{9771955}, massive MIMO-OTFS systems exploit the inherent 2D cluster structure in the Doppler-angle domain using the local Beta process and SBL, achieving superior performance and robust adaptation to variable Doppler conditions in comparison with the EM-VBL technique. In one of the earlier works in \cite{9483694}, a CSI estimation scheme using a BL framework for OTFS modulation systems has been proposed. Pilots are transmitted directly in the TF domain, reducing pilot load and training time, while estimation accuracy is improved by using DD domain sparsity.

In OTFS systems, fractional channel parameters reduce effective SNR and spread performance, particularly when the OTFS frame is small. As a result, it is critical to consider the Doppler effect and delay as fractions when designing the system. For instance, the study conducted in \cite{Wang_2023} examines the input-output relationship of the SISO-OTFS system using both integer grid and fractional grid. The fractional DD domain is sparser and more accurate than the integer DD domain. Furthermore, the work presents a model that includes both doubly fractional and mixed 1D and 2D fractional components. The hybrid concept provides a solution to the challenge of tandem off-grid managing. The proposed hybrid model approach combines off-grid decomposition with SBL to improve accuracy while lowering processing costs. In this approach, the received signal is broken down into two interconnected lattice structures and subsequently merged within the SBL framework. The results outperform conventional systems based on doubly fractional models in terms of accuracy and computing cost. CE with fractional Doppler can be considered as an off-grid sparse signal recovery problem where the virtual sampling grid is introduced in the DD domain. A traditional approach for estimating fractional Doppler and sparse signals in OTFS systems is the first-order linear approximation. However, the linear approach can lead to significant modeling errors in OTFS systems, as the finite sampling grid may not be accurate enough. To address this issue, another efficient SBL method \cite{Zhang_2024} is proposed to jointly estimate the fractional Doppler and sparse channel vectors. In particular, the input-output relation for fractional Doppler is reformulated, and a new off-grid CE model is proposed. Based on the off-grid sparse channel model, they presented a pilot design scheme. By analyzing the components of each received signal, they determined that IDI is distributed in the Doppler dimension. Therefore, accurate fractional Doppler estimation can be achieved in OTFS systems when pilots and guard intervals are arranged in the whole Doppler dimension. Then, the majorization-minimization (MM) algorithm is adopted to iteratively update the relevant parameters under the SBL framework. The fractional Doppler can be iteratively corrected to eliminate the off-grid gap using the gradient descent method. Expanding on the basis of SBL for OTFS-CE, recent progress has centered on overcoming the constraints of conventional on-grid techniques when confronted with fractional delay and Doppler parameters. Off-grid CE approaches have therefore emerged as a promising solution to mitigate the effects of channel spreading that degrade the inherent sparsity structure of the DD domain. Furthermore, study \cite{10506450} proposes an off-grid Bayesian compressive sensing (OG-BCS) algorithm and a fast variant (FBCS) using a Laplacian scale mixture prior to handle fractional DD parameters. This approach reduces computational complexity while achieving superior NMSE performance over traditional SBL and Laplace-based methods under varying pilot configurations and DD grid sizes. In \cite{9738478}, an off-grid OTFS-CE system is presented that uses a SBL framework to avoid channel spreading caused by fractional delay and Doppler shifts. This method formulates the problem as 1D off-grid sparse signal recovery based on a virtual sampling grid. It separates the on-grid and off-grid components using hyperparameters that are estimated via the EM algorithm. This achieves superior performance compared to traditional on-grid methods. Recently, a fast Bayesian compressive sensing (FBCS) method that integrates adaptive second-order grid refinement (ASGR) has been proposed for the estimation of doubly fractional channels in OTFS systems~\cite{11072479}. The ASGR algorithm employs FBCS for efficient hyperparameter estimation and employs a second-order Taylor expansion to improve modeling accuracy. The approach improves estimation accuracy and obtains superior NMSE performance in comparison to previous off-grid and CS techniques by iteratively refining the DD grid, all while maintaining computational efficiency. The development of off-grid estimation techniques is a significant advancement in OTFS-CE, particularly in scenarios where the actual channel parameters do not perfectly align with the discrete DD grid. This preserves the sparse structure that conventional OTFS algorithms rely on for optimal performance. To address the IDI problem caused by fractional Doppler, study in \cite{10038838} proposes a joint time and DD-CE method using SBL. The study in \cite{9184852} provides an SBL-based strategy that converts the CE to sparse signal recovery utilizing hierarchical Laplace prior and the EM algorithm. It demonstrates improved pilot overhead efficiency and NMSE performance compared to standard GP-CE and OMP-CE approaches. In \cite{9880804}, using efficient SBL (ESBL) with Doppler grid segmentation factor achieves enhanced computational efficiency and sparsity characteristics compared to traditional SBL-based algorithms.  In practical OTFS implementations, the Doppler squint effect poses additional issues that have been addressed by specific off-grid estimate approaches \cite{10475894}. A SBL-based approach specifically designed for OTFS systems with the Doppler squint effect employs EP patterns and off-grid SBL framework to recover DD channel parameters while accounting for subcarrier-dependent Doppler shifts. Among these, a recent study \cite{10677422} proposes a 2D off-grid CE method based on equidistributed information quantity, where grid evolution is guided by the estimated delay and Doppler path distribution. This technique adaptively refines the non-uniform grid structure.

Recent advances have positioned grid evolution-based techniques among the most comprehensive solutions to off-grid estimation challenges \cite{10436557}. In particular, the truncated grid evolution-based efficient sparse Bayesian inference (T-GEESBI) algorithm introduces adaptive grid-evolution mechanisms that iteratively refine the virtual DD grid instead of using fixed uniform grids, thereby achieving a superior accuracy–complexity balance while providing theoretical convergence guarantees for both ideal and practical rectangular-pulse scenarios. In \cite{10704024}, an adaptive grid refinement strategy is presented to address the issues of doubly fractional CE, contributing to the growing emphasis on flexible grid structures for CE. Using division and adjustment processes, this method dynamically refines the estimation grid, achieving superior performance with lower computational complexity and faster convergence than traditional uniform grid-based approaches. A recent study \cite{10758799} addresses the issues of CE in practical large-scale OTFS systems with the hardware constraint of low-resolution analog-to-digital converters (ADCs). An effective sparse CE framework based on VBL has been proposed for quantized MIMO-OTFS and orthogonal time sequency multiplexing (OTSM) systems with finite-resolution ADCs. The method employs an EM-based SBL algorithm with variational inference and incorporates a modified MMSE detector to address quantization effects by formulating the estimation as a quantized sparse recovery problem. This framework accomplishes favorable NMSE and symbol error rate (SER) performance in large-scale MIMO scenarios and provides robust CE under a variety of quantization resolutions.

\subsection{Matching Pursuit-Based Techniques}
\label{subsec:matching_pursuit}
MP techniques represent a fundamental class of greedy sparse recovery approaches that have gained increasing importance in CE tasks for wireless communication systems. Signals are iteratively decomposed into a linear combination of selected elements, commonly referred to as atoms, from an over-complete dictionary by these algorithms.  MP algorithms are particularly effective at exploiting the inherent sparsity observed in wireless communication channels due to this characteristic.

The sequential and greedy approach to signal reconstruction is the fundamental principle of MP algorithms. At each iteration, the algorithm identifies the dictionary element with the strongest correlation to the current residual signal. Subtracting the selected atom's contribution from the residual allows the algorithm to refine the approximation. The iterative refining process concludes when the predetermined ending criteria are met. These criteria include bringing the residual signal below a certain threshold, achieving an acceptable approximation accuracy, and completing the maximum number of iterations allowed.

MP algorithms are particularly well-suited for real-time applications in dynamic wireless environments due to their advantageous compromise between computational complexity and reconstruction accuracy.  Several extensions and enhancements, including orthogonal MP (OMP) \cite{8727425} and compressive sampling MP (CSMP) \cite{NEEDELL_2009}, have been developed to improve MP performance by addressing constraints like computational efficiency and convergence rate. This has increased MP's effectiveness and applicability in practical scenarios.

The study in \cite{8727425} addresses the challenging problem of downlink CE in OTFS massive MIMO systems, where the large number of base station antennas makes traditional estimation techniques computationally prohibitive. The key insight is that OTFS massive MIMO channels have a 3D structured sparsity pattern in the delay, Doppler, and angular dimensions.
To take advantage of this built-in sparsity structure, the authors suggest a 3D-Structured OMP (3D-SOMP) algorithm. The method sequentially estimates the support of each propagation path by formulating the CE as a sparse signal recovery problem. To reduce the effects of pilot interference, the approach uses random complex Gaussian pilot sequences with guard intervals. To address ISI challenges in massive MIMO-OTFS systems, study \cite{10213984} proposes a CE method using 3D inner product proportion reduce difference structured OMP (3D-IPRDSOMP) when delay path differences are smaller than system resolution. This method improves estimation accuracy by backtracking. The study in \cite{9128497} addresses the DD-CE in uplink OTFS multiple access systems. CS-based algorithms with OMP and MSP are proposed, which outperform impulse-based methods in terms of NMSE and BER. The work in \cite{9200896} addresses the computational complexity challenge in mmWave massive MIMO-OTFS-CE by proposing a tensor-based OMP algorithm with effective pilot design in the TF domain and exploiting sparsity in the delay-Doppler-angle (DDA) domain. 

OMP-based CE appears not only in unidirectional systems but also in uplink-downlink CE problems. As demonstrated in this study \cite{10001420}, in high-mobility scenarios, the use of information obtained from uplink CE for downlink becomes critical due to processing delay. The OMP algorithm provides effective results in both uplink and downlink CE by exploiting sparsity characteristics. However, as this study shows, in real-time applications, channel reciprocity can be degraded due to processing latency, significantly affecting downlink CE performance.

One of earlier work for CS-based CE in \cite{9685856}, CE scheme employing a continuous DD dictionary has been proposed, eliminating the assumption that OTFS channel delays are restricted to integer multiples of the sampling period. This approach significantly reduces estimation error compared to conventional integer-delay-based methods. In \cite{9590508}, an effective sparse CSI estimation model is proposed for MIMO-OTFS systems that can handle fractional Doppler effects and exploit simultaneous row-group (RG) sparsity, reducing pilot overhead and training duration while increasing spectral efficiency. In \cite{9891774}, a CSI estimation model is developed for MIMO-OTFS systems that exploits 4D-sparsity in DDA domain using the OMP framework. This achieves flexible pilot placement with significantly reduced pilot overhead and low training requirements. A study \cite{10201832} proposes a CS-based algorithm for MIMO-OTFS-CE that exploits the structured sparsity of DD domain channels using row-block orthogonal matching pursuit (RBOMP). The problem is formulated as RB sparse recovery and achieves superior performance compared to conventional methods.

\subsection{Message Passing-Based Techniques}
\label{subsec:message_passing}
A promising approach for OTFS-CE is offered by MEP-based techniques by exploiting the sparse structure of DD channels and enabling iterative information exchange between variable and factor nodes to achieve near-optimal estimation performance with reduced computational complexity. Building upon this approach, a MEP algorithm is developed in \cite{9456029} for OTFS-CE, which addresses fractional Doppler shifts through Bayesian-based structured sparse signal recovery. Similarly, for OTFS-UWA systems, a generalized approximate message passing (GAMEP) method based on structured sparsity was developed, using the structured sparsity features of doubly spread channels with lower computational complexity \cite{Zhang_2023}. The method provided in \cite{10310071} uses structured sparsity in the DD domain effective channel under fractional Doppler conditions and is based on a hidden Markov model (HMM). This method uses EM and unitary approximate message passing (UAMEP) to optimize hyperparameters. By using an efficient factor graph representation, the proposed scheme achieves notable performance gains over conventional SBL techniques. In \cite{9864300}, a joint CE and SD framework using MEP has been developed for hybrid RIS (HRIS)-aided mmWave OTFS systems. Furthermore, block time-varying CE with limited pilots has been proposed \cite{10870378}, which employs ML and extended Kalman filter (EKF)-based algorithms to jointly estimate dynamic channel parameters in OTFS systems. This approach achieves a favorable MSE and complexity performance in comparison to conventional methods.
\subsection{Deep Learning-Based Techniques}
\label{subsec:deep_learning}
Due to their dependence on exact statistical models and prior knowledge, conventional model-based CE techniques typically do not accurately represent the dynamic and complex nature of wireless channels. However, DL has become a compelling data-driven solution for OTFS-CE since it can characterize highly nonlinear interactions between pilot signals and channel properties without requiring explicit analytical formulations. Deep neural networks (DNNs) can extract salient features from received signals and generalize across a wide range of channel conditions because they learn directly from data. Architectures such as convolutional neural networks (CNNs) and recurrent neural networks (RNNs) have demonstrated notable success in capturing the channel's underlying DD structure, providing high estimation accuracy and computational efficiency suitable for real-time deployment.

In \cite{9896637}, DL-CE technique is proposed for OTFS systems, which uses classic OMP algorithm for preliminary estimation and residual network (ResNet) to refine results. This achieves significantly better performance than conventional methods with good robustness across different scenarios. In \cite{10138432}, a DNN-based fractional Doppler CE scheme is proposed in air-to-ground communication scenarios, using received pilots as network inputs to estimate channel parameters and achieving higher accuracy with lower pilot power consumption through a simple two-hidden-layer DNN architecture. Additionally, by utilizing the advantageous training properties of TF domain matrices in contrast to DD domain, TF domain learning approaches \cite{10439989} demonstrated significant computational advantages, achieving comparable estimation performance with much lower complexity for practical uses.  

DL-based approaches have emerged as effective solutions for OTFS-UWA-CE challenges \cite{10400942}. A learned denoising method employs FastDVDNet, originally designed for video denoising, to treat multiple DD domain channel estimates as correlated noisy images derived from the same clean reference. By intelligently utilizing inter-estimate correlation patterns, this DL framework achieves better NMSE and BER performance than conventional sparse estimating algorithms in time-varying UWA scenarios. In addition to denoising-based and TF-domain learning frameworks, \cite{10540188} introduces a DL-based CE approach that explicitly addresses nonlinear transmission impairments caused by high-power amplifiers (HPAs). The proposed method uses a long short-term memory (LSTM) neural network (NN) operating in the TF domain, which outperforms conventional DD-domain estimators in terms of BER, PAPR, and throughput under highly dynamic vehicular communication scenarios. The approach outperforms traditional and learning-based estimators in terms of NMSE and failure rate performance while maintaining reasonable computational costs owing to the incorporation of a learnable proximal mapping layer that makes use of sparse priors and Bayesian regularization principles. Using a deep residual attention network (DRAN) to improve CE accuracy in OTFS systems, another effective method \cite{10856399} extracts pilot information while suppressing data interference. In \cite{10747184}, a 3-phase scheme using an autoencoder and RNN predicts channel variations in high-mobility scenarios, achieving superior accuracy in CSI element extraction with linear computational complexity. A model-driven CE network (CENet) has been proposed for OTFS systems operating in the delay-time (DT) domain in another DL-based study employing a different architectural approach \cite{10916513}. The approach utilizes initial least-squares (LS) solutions as prior information for training, employing a LS-driven CENet with parallel structure to efficiently handle high-dimensional DT domain channel matrices. 

Recent research has provided additional evidence that DL designs are flexible and effective in complicated, cooperative communication situations. \cite{10994238} uses DL-based models (ECSIEst-Net) and data-driven networks (SSymDet-Net) to detect CE and secondary symbols in symbiotic radio OTFS systems. In the DD domain, these networks benefit on highly structured Doppler shifts and combinatorial frequency modulations by using CNN architectures. DL-based approaches have also been effectively applied to CE and SD in RIS-OTFS systems. In \cite{11010104}, a DL-based secondary information detector (SecNet) and a DL-SBL estimator are proposed for RIS-assisted symbiotic radio OTFS systems. These methods demonstrate significant performance gains over conventional methods. A recent advancement presents a lightweight DL-based end-to-end joint CE, data detection, and demodulation (JCEDD) framework for OTFS systems \cite{11071963}. This method uses data augmentation techniques to achieve better performance in NMSE and BER while keeping model complexity low. Compared to traditional JCEDD and pilot-based methods, it provides a more efficient and scalable alternative for realistic OTFS implementations.

\subsection{Recent Techniques}
\label{subsec:recent_techniques}
Recent CE algorithms are critical for addressing the limitations of traditional methods in next-generation wireless systems characterized by high-mobility, massive MIMO, and mmWave communications.
In \cite{10122554}, low complexity for CE in single-antenna OTFS systems is proposed. This method utilizes fine DD resolution to decouple the joint estimation of channel parameters into separate delay and Doppler estimations for each path, resulting in superior accuracy without the need for matrix inversion in comparison to conventional methods. \cite{10666708} advances the low-complexity CE paradigm by presenting an effective single-tone (ST)-based estimate framework that successively derives channel parameters through threshold-based detection. This approach, which is specifically designed to handle both integer and fractional Doppler shifts, introduces a refined parameter extraction mechanism that allows for better accuracy-complexity trade-offs, contributing significantly to the development of efficient and scalable OTFS-CE techniques. \cite{10557727} presents a weighted LS (WLS) channel parameter estimate method for OTFS systems with fractional Doppler shifts. This method reduces fractional Doppler effects by using an iterative interference suppression technique along with the characteristics of block sparsity in the DD domain. A method called 3D efficient subspace pursuit (3D-ESP) for MIMO-OTFS systems is introduced in \cite{Gui_2024}, utilizing the unique features of the limited DDA domain. A noise-adaptive filtering criterion is included in the method to improve stability in noisy conditions, and LS optimization is applied to identify several relevant paths in each iteration. The Matrix Pencil approach is proposed as a CE strategy for OTFS systems with fractional Doppler shifts in \cite{10582886}. Using IDFT, the DD domain problem is restructured as an estimation of exponential sinusoid parameters in the DT domain, converting Doppler shifts and channel gains into frequency and amplitude parameters. The algorithm decomposes coherent Doppler shifts without requiring Doppler search space. In \cite{Guo_2024}, a CS-basis expansion model (CS-BEM) is proposed to minimize pilot overhead while maintaining estimation accuracy. The technique allows for the simultaneous extraction of channel sparse information and spread information in the Doppler domain by deriving a complex exponential BEM in the DD domain. The study \cite{10612829} introduces a structured sparsity adaptive MP (SSAMP) algorithm for MIMO-OTFS-CE, tackling the issue of unknown channel sparsity in practical scenarios. The method uses a flexible step size approach to work efficiently, even when the channel sparsity is unknown. It includes different step sizes to speed up reconstruction and backtracking techniques to enhance accuracy. A modified SSAMP (MSSAMP) strategy is proposed, which demonstrates better NMSE performance than traditional OMP and SOMP methods, while also enhancing reconstruction time and maintaining similar estimation accuracy. Two choice hard thresholding pursuit (TCHTP), a low complexity fast greedy sparse recovery technique, is developed in \cite{10073950} to estimate the number of CSI and unknown DD paths while outperforming state-of-the-art (SOTA) methods with significantly reduced computational complexity. 

For OTFS-based applications in ISAC systems, a DD inter-path interference cancellation (DDIPIC) algorithm has been designed to enable joint CE and radar parameter estimation with superior range and velocity accuracy \cite{Muppaneni_2023}. Using corner-inserted pilot pattern (CIPP) and LMMSE estimation, study \cite{9400868} proposes a transform-domain basis functions technique for low-dimensional subspace channel modeling in OTFS systems, addressing the complexity issues from a subspace modeling perspective. The method significantly reduces computational complexity while achieving superior MSE and BER performance compared to conventional SI-based and CS-based CE methods in high-mobility scenarios. Additionally, study \cite{10385066} offers a CE approach based on the discrete Zak transform (DZT) that decouples the estimate of delay, Doppler, and channel coefficients using structured matrix decomposition. This scheme achieves low complexity while exhibiting superior NMSE and BER performance in a variety of mobility scenarios. In overspread channels, where channel delay spread exceeds block duration, specialized two-stage estimation approaches have been developed in \cite{Priya_2024}. This method first uses a DD domain EP scheme to estimate aliased delays and Doppler shifts, followed by time domain dual chirp correlation in the second stage to estimate actual delays and Doppler shifts for remaining paths. The approach addresses the challenges of overspread channels by resolving ambiguity in estimating delays and Doppler shifts for paths sharing the same aliased delay. 

Another line of research addresses OTFS-based CE by considering synchronization aspects. In particular, study \cite{10742117} proposes a pilot-aided technique that uses maximum length sequences (MLS) to do both time synchronization and off-grid Doppler estimates at the same time. By combining synchronization and CE into one framework, the proposed technique improves estimation accuracy. Another effective technique \cite{10816508} is differential modulation-aided CE with lightweight network assistance, which enables pilot-free operation while reducing complexity and enhancing spectral efficiency in uplink OTFS systems. An efficient smoothed L0 (SL0) algorithm \cite{10938192} has been proposed for OTFS-CE in scenarios where channel sparsity is unknown. This CS approach defines CE as sparse signal recovery, which eliminates the need for prior sparsity knowledge. This approach allows for rapid and precise estimation in uncertain channel conditions, while simultaneously balancing computational complexity and accuracy. For high-mobility OTFS systems, a recently proposed method \cite{10993437} extends the boundary of CE. A pilot-aided CE algorithm based on the modified discrete Chirp Fourier transform (MDCFT) is developed, specifically designed for highly dynamic scenarios with rapidly varying Doppler shifts. The method models the phase of the pilot channel impulse response in the DD domain as a second-order polynomial and leverages the Zak transform to establish an accurate input-output relationship. 

\section{Joint Channel Estimation and Signal Detection}
\label{sec:joint_estimation}
Traditional CE systems typically perform CE and SD as independent, sequential procedures. Joint techniques, which execute CE and data detection at the same time, provide superior results by capitalizing on the interdependence of both tasks. One of the most comprehensive contributions in this domain is presented in \cite{10138088}, which develops an innovative orthogonal affine-precoded superimposed pilot (AP-SP) framework for both SISO and MIMO-OTFS systems. This framework presents two different BL methods: one is a pilot-aided technique that uses EM for CSI estimation, and the other is a data-aided approach that combines CE and data detection while considering uncertainty in channel state information during the detection process. The proposed architecture leverages arbitrary Tx-Rx pulse shaping and implements affine precoding within the DD domain, establishing theoretical Bayesian CRLB and demonstrating significant performance improvements over conventional techniques in high-mobility environments. Similarly, \cite{10147252} provides the first complete comparison between OTFS and OFDM systems under practical channel coding conditions, presenting a BEM-CE approach that dramatically decreases pilot power while increasing estimation accuracy. This study demonstrates that OTFS outperforms OFDM when utilizing small modulations such as BPSK and QPSK with the proposed BEM-based channel estimator, highlighting the potential of coded OTFS systems for future wireless communications. A joint CE-SD study suggests a CS-based approach that combines iterative processing with pilot optimization, using the BEM \cite{10887291}. This approach demonstrates how BEM-based frameworks improve estimate and detection performance in OTFS systems while maintaining computational efficieny. The study \cite{9785832} presents a hybrid CE and SD strategy for OTFS-based LEO satellite communications, utilizing the VB inference (VBI) architecture. Hidden variables that carry unknown data symbols and act as virtual pilots improve CE's accuracy and data symbol recognition. This produces superior NMSE and BER results than standard techniques. SP schemes have developed as an effective way for combining CE and SD in OTFS systems, providing higher spectral efficiency than conventional pilot-aided methods \cite{10475591}. The SP-DD receiver uses BEM and generalized complex exponential BEM for channels. It uses MEP-based iterative processing to perform both estimation and detection at the same time. This approach strategically designs pilot signals to achieve pilot power concentration in the frequency domain, facilitating the reduction of pilot power without decreasing the performance of the receiver while simultaneously reducing PAPR of OTFS signals. In MIMO-OTFS systems, the challenges posed by fractional Doppler effects and high computational complexity have been effectively addressed through advanced joint processing techniques \cite{10476726}. One significant contribution is the combination of joint CE and symbol SD algorithms with adaptive block OMP (ABOMP), which is made possible by a partially iterative feedback mechanism that dynamically connects the CE and SD modules to improve system performance. 

According to \cite{Yang_2024}, iterative path peak search (IPPS) algorithms effectively address fractional DD cases. Using the relationship of path responses to find the first peak positions, the IPPS method redefines CE as a path peak-seeking problem in the DD domain. These estimates are then iteratively refined using the limited-memory Broyden–Fletcher–Goldfarb–Shanno (L-BFGS) optimization method. To improve CE accuracy and reduce inter-path interference, the IPPS algorithm optimizes pilot responses while simultaneously refining the detected paths. In \cite{9864300}, a proposed joint CE and data detection (JCEDD) scheme for HRIS-aided mmWave OTFS systems uses an EM algorithm to accurately estimate channel parameters and detect data symbols. Another recent study \cite{9928043} designed joint successive activity type identification, CE, and SD methods for LEO satellite-based Internet of Things (IoT) networks within the GF-NOMA framework. Using location-aided CE and iterative interference cancellation (IIC) detection, \cite{RISOTFS14} optimizes OTFS frame structure and RIS phase shifts simultaneously, leading to significant performance gains in low-complexity, high-mobility environments. Recent research suggests that parametric bilinear Gaussian belief propagation (PBiGaBP) approaches are effective joint estimation methodologies \cite{10683792}. These techniques improve the estimation accuracy by reducing OTFS demodulation to a straightforward parametric bilinear inference problem. PBiGaBP uses a simple step-by-step approach to find an adequate balance between performance and efficiency, making it a great choice for practical OTFS applications. Recently, the joint estimation and detection problem has also been addressed for clipped OTFS systems \cite{11072043} under nonlinear power amplifier distortions. In this framework, a SBL-based joint channel and clipping amplitude estimator (SBL-JE) is integrated with Kalman filtering and MMSE-based decision feedback blockwise equalization (DFBE), enabling iterative tracking and robust SD.

\section{Integration of OTFS Channel Estimation with Emerging Technologies}
\label{sec:next_gen_systems}
In the context of CE, the combination of OTFS modulation with emerging technologies like ISAC, RIS, mmWave communication, massive MIMO, NOMA, IoT, LEO satellite systems, and UAV communications presents both opportunities and challenges. Utilizing the enhanced freedom and sensing capabilities offered by these advanced technology necessitates resolving challenges related to DD domain processing.

In mmWave MIMO systems, BL methods exploit DD sparsity achieving superior CE performance \cite{9827947}. For asymmetrical massive MIMO systems with different numbers of transceiver radio frequency chains, \cite{10272673} introduces a new two-stage estimation method that uses coprime patterns and virtual array reconstruction. This method is capable of reducing into the 3D DDA search area to two dimensions, improving CE accuracy and lowering computational complexity compared to traditional 3D search methods. In \cite{10472131}, joint sparsity pattern learning techniques have been developed for massive MIMO-OTFS systems to exploit the inherent sparsity characteristics across multiple antennas by employing flexible spike and slab prior models in the DDA domain. For massive MIMO-OTFS systems, addressing high-dimensional sparse representation challenges, a fast sparse recovery approach has been proposed using practical rectangular transmitted waveforms \cite{10915571}. The work \cite{10690180} presents an adaptive block sparse backtracking technique that modifies block sizes according to the features of the residual signal. When compared to COMP techniques in massive MIMO-OTFS systems, this approach achieves up to a 20$\%$ reduction in estimation error, greatly improving estimation performance. A comprehensive framework integrating RIS and mmWave technologies has been developed for HRIS-aided OTFS systems, employing joint CE and SD approaches that exploit DD sparsity \cite{9864300}. In addition, RIS-aided OTFS systems use efficient uplink transmission techniques with sensing capabilities for high-mobility scenarios, utilizing EKF for channel tracking \cite{10100890}. In another RIS-OTFS system, a comprehensive 2D off-grid sparse CE model has been developed to address fractional Doppler effects in multi-frame scenarios \cite{10976498}. The approach exploits DD sparsity in cascaded RIS-OTFS channels, formulating CE as a block sparse recovery problem using B-CPSBL framework while proposing low-complexity phase optimization algorithms to maximize achievable rates. 

Addressing the challenges of off-grid parameters in mmWave systems, \cite{10164149} presents a cross-domain CE scheme for OTFS-based mmWave hybrid beamforming systems that establishes appropriate signal models in both time domain and DD domain. The approach can convert the 2D channel parameter estimation problem into parallel sparse recovery problems, which are solved through an SBL-based BEM algorithm. To deal with nonlinear problems in mmWave systems, threshold-based CE has been designed for OTFS systems with PA distortions. It is more resilient than traditional linear approaches \cite{10100890}. 

In ISAC systems, OTFS modulation allows for simultaneous exploitation of sensing parameters for improved communication performance \cite{10409528}. An efficient joint parameter association, CE and SD (PACESD) framework addresses the challenges of sensing-aided vehicular networks by formulating the problem as a constrained bilinear recovery where sensing parameters acquired through roadside unit downlink transmission are leveraged to improve uplink performance. The approach employs a bilinear unitary approximate MEP (Bi-UAMEP) algorithm to solve the joint estimation problem. In another OTFS-ISAC work \cite{Muppaneni_2023}, OTFS-CE employs DD inter-path interference cancellation algorithms for joint CE and radar parameter estimation. In the study \cite{10714398}, an OTFS-based architecture is used to improve ISAC systems in environmental sensing by simultaneously estimating target state and transmitter location. The approach employs a WLS formulation enhanced by semi-definite relaxation (SDR) techniques, enabling accurate sensing performance. An advanced OTFS-ISAC framework that makes use of both BL and BS learning techniques is proposed in \cite{11053775}. By exploiting DD-domain sparsity, this framework can jointly estimate radar target parameters and wireless channel coefficients. The framework demonstrates near-BCRLB NMSE performance and robustness against multipath and fractional Doppler effects. Beyond simulation-based evaluations, the practical implementation and validation of OTFS-ISAC systems have recently been demonstrated on software-defined radio testbeds. In particular, a low-complexity CE framework that integrates parameter-inherited SBL (PI-SBL) and UAMEP has been developed and experimentally validated in high-mobility V2X environments. By utilizing large-scale matrix operations via 2D FFT, the suggested technique offers robust and efficient joint communication and sensing performance. Furthermore, robust DD pilot structures and $\ell_0$-MGMVC-based CE have been recently introduced for OTFS-ISAC systems~\cite{11058951}, effectively addressing high PAPR and providing outlier-robust estimation in both communication and sensing tasks. Furthermore, for OTFS-ISAC, an effective sequential path estimation algorithm that takes advantage of energy leakage for inter-path interference elimination has been proposed \cite{11068135}, providing better estimation accuracy and scalability under fractional effects with reduced computational complexity. For OTFS-based multiuser scenarios in NOMA systems, \cite{10540130} suggests a cross-domain serial interference cancellation (CD-SIC) technique.The technique reconstructs the DD domain channel response through an iterative refinement process after performing LS-CE in the TF domain. The suggested framework outperforms traditional SIC schemes by combining this method with MMSE equalization, which greatly enhances BER performance, user fairness, and estimation robustness under changing mobility and power allocation conditions. In \cite{fd_otfs}, a two-step OMP with fractional refinement technique is suggested for SISO radar in OTFS-based ISAC systems, effectively capturing both integer and fractional DD with significantly reduced complexity compared to classic OMP by avoiding large dictionary matrices. For LEO satellite-based IoT networks, successive CE methods within GF-NOMA paradigm demonstrate enhanced performance \cite{9928043}. A framework based on generative adversarial networks (GANs) \cite{10815946} has been developed to address CE in OTFS-UAV systems in high-mobility conditions. To facilitate OTFS system operation in high-mobility settings, a GAN-based CE framework for UAV communications has been developed. The proposed model can achieve reliable CE by employing a U-Net generator and a PatchGAN discriminator. Its performance has been validated against traditional estimators under a variety of modulation schemes and velocities, thereby demonstrating its suitability for real-time UAV applications.
\section{OTFS Channel Estimation Challenges}
\label{sec:OTFS_Challenges}
Although OTFS modulation offers considerable benefits in high-mobility channel conditions, the practical use of channel estimation techniques faces notable difficulties. These challenges stem primarily from hardware constraints and the fundamental properties of real-world wireless channels, which deviate from theoretical idealizations. This gap between theory and practice presents a critical barrier that must be addressed. To fully realize the promise of OTFS systems, various restrictions in theoretical modeling and actual implementation need to be discussed.

Signal processing encounters various technical challenges in practical OTFS implementations. The primary concerns are managing leakage from fractional DD parameters, controlling IDI in fractional cases, and ensuring robustness in impulsive noise environments. Implementation challenges include signaling overhead optimization, guard space requirements, PAPR management, beam squint effects in wideband systems, and hardware limitations. These signal processing and implementation issues are frequently interconnected, necessitating detailed design approaches. Each of these challenges necessitates specialized solutions and can be an active area of ongoing research in the OTFS community. Understanding and addressing these challenges enables OTFS-based systems to be successfully deployed in practical wireless communication networks.
\label{subsec:challenges}
\subsection{Leakage Suppression}
\label{subsec:leakage_suppression}
Leakage effects can be a significant challenge in the real DD domain, OTFS-CE. These effects occur when the channel characteristics do not fully match the discrete DD grid. As a result, energy flows across multiple grid locations, affecting the sparsity structure upon which OTFS-CE algorithms rely. This leakage effect, especially in situations involving fractional delays and Doppler shifts, significantly reduces measurement precision and estimation accuracy. \cite{9737331} discusses a CE technique with leakage suppression for OTFS modulation to address this critical issue. The scheme uses smoothness regularization in the TF domain to reduce leakage effects in the DD domain. The method improves CE accuracy while reducing signaling overhead, resulting in better performance in linear time-varying channel scenarios without periodicity assumptions.
\subsection{Inter-Doppler Interference}
\label{subsec:inter_doppler_interference}
OTFS systems with fractional Doppler shifts display IDI, leaking energy between Doppler bins and reducing CE performance. This interference changes the position and height of response peaks in the DD domain, making precise parameter determination difficult. A two-stage algorithm addresses this issue by first performing rough position estimation of nonzero Doppler shifts and then employing the quasi-Newton method for iterative re-estimation, achieving superior performance than conventional correlation-based and threshold-based approaches \cite{He2023ATC}. A VRCF CE approach \cite{11023080} has been proposed to reduce IDI in fractional Doppler conditions. This method improves on conventional strategies in terms of NMSE and BER by integrating multi-stage Doppler estimates with adaptive interference cancellation.
\subsection{Impulsive Noise}
\label{subsec:impulsive_noise}
Impulsive noise poses a significant challenge to OTFS-CE, as it deviates from the assumptions of conventional AWGN in practical deployment scenarios. Unlike Gaussian noise, impulsive disturbances are characterized by random, high-amplitude interference that can severely distort the sparse structure of the DD domain. This structure is fundamental for accurately estimating the parameters of the channel in OTFS systems. In order to overcome these limitations, robust estimation methods have been developed, such as $l_0$-norm constrained maximum Versoria criterion (MVC) technique \cite{10209369}, which show better resilience against impulsive noise than traditional LS methods while preserving the benefits of OTFS modulation.
\subsection{Signalling Overhead}
\label{subsec:signalling_overhead}
Signalling overhead can be a significant difficulty in OTFS-CE, as accurate DD domain CSI requires the transmission of pilot symbols, which reduces spectral efficiency. To address these limitations, advanced 1D CE techniques have been introduced, including frequency-domain pilot-aided OTFS (FD-PA-OTFS), time-domain pilot-aided OTFS (TD-PA-OTFS), and time-domain training sequence-based OTFS (TD-TS-OTFS) variants \cite{10223427}. Among the proposed methods, FD-PA-OTFS achieves the most favorable balance, delivering a minimal signaling overhead of just 5.4$\%$ while maintaining enhanced diversity performance and robust CE accuracy. According to \cite{10356117}, online SBL (OSBL) frameworks can improve CE performance while significantly lowering pilot overhead compared to traditional EP techniques. Furthermore, new 2D pilot design approaches have been developed for multi-antenna OTFS systems. In \cite{10370744}, an efficient 2D pilot scheme is proposed that places pilots from different antennas in the same DD domain region using code division multiplexing (CDM) to mitigate inter-antenna pilot interference, achieving significantly better NMSE performance with lower pilot overhead compared to conventional single-pulse and sequence pilot schemes. In \cite{10587305}, a user-specific CE overhead optimization and resource allocation scheme is proposed for multi-user OTFS systems to address inter-grid interference (IGI) challenges. The method exploits user-specific statistical channel characteristics by dividing the frequency band into multiple sub-bands for multiple access, allowing individual optimization of each user's overhead in the corresponding DD domain. An alternating optimization algorithm with closed-form solutions is designed to jointly optimize overhead and bandwidth allocation to solve the non-convex capacity maximization problem. This algorithm balances the trade-off between IGI and transmission efficiency for increased spectral efficiency.
\subsection{Guard Space Requirement}
\label{subsec:guard_space}
Guard space allocation is a significant challenge in OTFS-CE schemes, particularly in joint estimation and detection frameworks where both pilot and data symbols are embedded in the same transmission frame. In order to minimize pilot-data interference, traditional methods necessitate the addition of guard intervals around pilot symbols, which significantly reduces spectral efficiency and increases signaling overhead. Recent advances have addressed this issue by combining novel SP-data placement strategies with advanced signal processing techniques. A method based on mean-field orthogonal approximate message passing (MF-OAMEP) has been shown to perform joint CE and data detection without guard space requirements \cite{10264119}. Iterative information exchange between channel estimators and symbol detectors is used to mitigate interference and maintain robust performance. In addition to MF-OAMP, a complementary strategy is to optimize the pilot structure and pulse shaping to reduce interference without requiring large guard regions. To this end, study \cite{10637960} proposes a CE scheme tailored for EP-based OTFS-CE systems, which incorporates non-rectangular TF-domain windowing (e.g., Slepian and Kaiser) and a particle swarm optimization (PSO)-based pilot sequence optimization algorithm. This approach significantly tightens the CRLB and improves both NMSE and BER performance in scenarios with severe data interference, all while maintaining a compact pilot structure and avoiding excessive guard space. 
\subsection{PAPR}
\label{subsec:PAPR}
The high PAPR in OTFS-CE remains a critical concern, particularly in systems that employ SP structures. In practical transmission scenarios, elevated PAPR not only reduces power efficiency due to the nonlinearity of power amplifiers, but it also degrades overall system performance. Traditional SP-based pilot schemes, such as superimposed single pilot (SSIP) and sparse pilot (SSPP), frequently require high-power pilot symbols to ensure accurate CE. This causes the PAPR issue and has an adverse effect on the system's spectral and energy efficiency.

In order to address such challenges, an efficient two-stage iterative CE (TSICE) algorithm is introduced in \cite{10550432}. This algorithm employs a hybrid SP (HSP) structure that combines a high-power center pilot and low-power edge pilots. In the initial stage, this design enables the estimation of DD taps through the use of the center pilot. Subsequently, the multi-path gain is refined using MMSE in the second stage. The method not only accomplishes superior BER performance in comparison to conventional SSIP and SSPP schemes but also mitigates PAPR by avoiding uniformly high pilot power allocation, all while eradicating the necessity for threshold tuning. In \cite{10614842}, a zero-bin low-PAPR (ZBLP) pilot design is proposed for zero-pad OTFS systems that strategically inserts Zadoff-Chu (ZC) pilot sequences into zero bins, effectively reducing both overhead and PAPR by spreading pilot symbols within zero bins while maintaining CE accuracy through a two-step estimation approach with low-complexity maximum ratio combining (MRC) detection. Another recent technique \cite{11030215} leverages ZC sequence-based pilot design to achieve both low PAPR and robust CE in OTFS systems. The proposed iterative ZC sequence-based estimation scheme enables accurate CE for both integer and fractional Doppler shifts while significantly lowering PAPR compared to conventional pilot arrangements.

\subsection{Beam Squint Effect}
\label{subsec:beam_squint}
Beam squint effects can pose a significant challenge to the performance of massive MIMO-OTFS systems operating in wideband millimeter-wave communications. The array response deviates from the intended direction due to beam squint, which is characterized by a change in beam direction across different frequencies within a wideband signal. In high-mobility scenarios, this effect gets worse due to Doppler squint, which causes further variations in beam direction due to Doppler shifts. The doubly squint effect, which is the result of the combination of these effects, can result in a significant decrease in the reliability of data transmission and the accuracy of CE.

An efficient hybrid precoding and CE scheme has been developed to mitigate the doubly squint effect in massive MIMO-OTFS systems that operate in high-mobility wideband scenarios \cite{10970022}. The method introduces a comprehensive system model that is based on the doubly-squint effect. It derives input-output relationships in the DD domain and proposes a chirp pilot-based CE structure that incorporates both beam squint and Doppler squint effects. The doubly squint effect is mitigated by a hybrid precoding method that combines digital and analog precoding. Analog precoding addresses beam squint, while digital precoding addresses Doppler squint.

\subsection{Hardware Impairments} \label{subsec:hardware_impairments}
Wireless systems that rely on OTFS also face the significant challenge of hardware impairments. These limitations can significantly reduce CE accuracy and overall system reliability, particularly in practical deployments and high-mobility environments. Unlike idealized theoretical models, practical transceivers are susceptible to a variety of non-ideal characteristics such as amplifier nonlinearities, phase noise, I/Q imbalance, and quantization errors. As a result of these limitations, CE's BER and MSE increase, distorting the received signal and making accurate CSI estimation a challenge.

The adverse impact of hardware impairments on OTFS-CE has been examined in recent studies. For example, in the presence of hardware impairments, \cite{11014223} suggests a low-complexity BEM-based CE framework for OTFS. This method offers a theoretical characterization of the MSE and illustrates, through both analytical derivation and simulations, that the estimation and error performance of OTFS systems can be dramatically impacted by even moderate hardware impairments. 

Finally, the studies discussed in this work are summarized; the parameters used across these studies are detailed in Table \ref{tab:notation}, while the CE methodologies are outlined in Table \ref{tab:OranWorkgroups2}.
\section{Conclusion}
\label{sec:conclusion}
This survey focused on CE techniques for OTFS modulation, categorizing the methodologies into DD domain approaches, TF domain methods, and various algorithmic frameworks. The DD domain approach stands out particularly. The DD domain representation provides sparse channel modeling advantages for high-mobility scenarios, with separate pilot, EP, and SP techniques serving as the main research directions. Several algorithmic approaches have been discussed, including BL, MP, MEP, DL, and joint CE-SD strategies, each with distinct trade-offs in terms of accuracy, efficiency, and complexity. While these methods appear promising in theory, practical application reveals additional complexities. Implementation challenges such as leakage suppression, IDI mitigation, signaling overhead reduction, and hardware impairments have been identified, as has the integration of OTFS-CE into emerging technologies such as massive MIMO, mmWave communications, RIS, and ISAC. While ongoing research advances both algorithmic and implementation aspects, additional study is needed to address practical deployment challenges, particularly complexity-performance trade-offs and robustness under realistic conditions. This survey presents an overview of OTFS-CE approaches and can assist researchers understand the current state of the field, including its inherent challenges and limitations, particularly in the context of 6G and beyond wireless networks.

\begin{table}[!htbp]
\centering
\caption{Parameters used in complexity analysis (as shown in Table~\ref{tab:OranWorkgroups2})}

\textit{Note: \# denotes "number of"}
\renewcommand{\arraystretch}{1.3}
\resizebox{\columnwidth}{!}{%
\begin{tabular}{|p{1.8cm}|p{3cm}|p{1.8cm}|p{3cm}|}
\hline
\rowcolor{lightgray}
\textbf{Parameter} & \textbf{Description} & \textbf{Parameter} & \textbf{Description} \\
\hline \hline
$M$ & \# delay bins in OTFS grid & $N$ & \# Doppler bins in OTFS grid \\
\hline
$M_A$ & \# antennas & $M_{RA}$ & \# receive antennas \\
\hline
$M_{\mathrm{TA}}$ & \# transmit antennas & $N_{\mathrm{TS}}$ & \# training symbols \\
\hline
$P$ & \# propagation paths & $N_\mathrm{ite}$ & \# iterations \\
\hline
$N_P$ & \# pilot symbols & $N_F$ & \# pilot symbols in frequency \\
\hline
$N_T$ & \# pilot symbols in time & $G$ & \# angle grid \\
\hline
$N_{\mathrm{DO}}$ & \# pilot symbols in Doppler & $N_{\mathrm{DE}}$ & \# pilots symbols in delay \\
\hline
$G_{\mathrm{DO}}$ & guard length in Doppler & $G_{\mathrm{DE}}$ & guard length in delay \\
\hline
$M_\tau$ & \# delay bins (sparse grid) & $N_\nu$ & \# Doppler bins (sparse grid) \\
\hline
$\mathcal{M}$ & modulation order & $N_{\mathrm{out}}$ & \# outer iterations \\
\hline
$C_L$ & channel length & $M_\mathrm{dom}$ & \# dominant delay taps \\
\hline
$J$ & \# user & $k_\nu$ & maximum Doppler tap \\
\hline
$l_\tau$ & maximum delay tap & $\tilde{N}$ & \# neighboring Doppler interference term \\
\hline
$G_T$ & guard length in time & $N_{\mathrm{IoT}}$ & \# IoT terminals \\
\hline
$N_{\mathrm{act}}$ & \# identified active terminals & $N_{B}$ & \# bases vector \\
\hline
$n_b$ & \# digital bits of precision & $P_\mathrm{DE}$ & \# delay path \\
\hline
$\epsilon$ & estimation accuracy & $\Delta$ & search step size \\
\hline
$n_\nu$ & \# Doppler refinement points & $m_\tau$ & \# delay refinement points \\
\hline
$N_\mathrm{dic}$ & \# selected row from dictionary matrix & $\mathrm{Q}_{n}$ & \# neurons in $\mathrm{n}^{th}$ hidden layer \\
\hline
$N_\alpha$ & \# sampling grid points in delay & $N_\beta$ & \# sampling grid points in Doppler \\
\hline
$\Theta(MN, MN)$ & complexity of EVD of an $MN \times MN$ matrix & $M_0$ & \# truncated sample after windowing \\
\hline
$\hat{N}$ & \# fractional Doppler bins on each side of the integer Doppler support & $\Delta\alpha$ & step size in the delay \\
\hline
$\Delta\beta$ & step size in the Doppler & $Q_{\mathrm{BEM}}$ & BEM order \\
\hline
$\Psi$ & \# selected atoms & $N_e$ & \# non-zero channel elements \\
\hline
$\mathcal{B}$ & \# block size & $K_{\hat{\ell}}$ & \# Doppler indices for aliased delay $\hat{\ell}$ \\
\hline
$J_{\hat{\ell}}$ & \# Doppler indices for aliased delay $\hat{\ell}$ for second stage & $\lambda$ & \# historical gradients \\
\hline
$M_p$ & \# delay bins in CE region & $N_p$ & \# Doppler bins in CE region \\
\hline
$M_\mathrm{on}$ & \# active subcarrier & $M_d$ & \# data subcarrier \\
\hline
$k_\mathrm{min}$ & \# minimum Doppler tap & $\chi$ & stopping criteria \\
\hline
$L_\mathrm{DB}$ & \# layers in the denoising block & $c_n$ & \# channels in the n$^{th}$ layer \\
\hline
$s_n$ & filter size n$^{th}$ layer & $N_\mathrm{RF}$ & \# RF chains \\
\hline
$K_\mathrm{SE}$ & \# column of sensing matrix & $\alpha$ & scale parameter \\
\hline
$B_\mathrm{ED}$ & \# encoder-decoder block & $L_L$ & layer size \\
\hline
$N_\mathrm{conv}$ & \# convolution filters & $Q_F$ & \# frame \\
\hline
$N_a$ & \# accerelation searches & $N_\mathrm{GS}$ & \# Doppler grid searches \\
\hline
$g_k$ & width of the guard cells in Doppler & $r_k$ & width of the reference cells in Doppler \\
\hline
$g_l$ & width of the guard cells in delay & $r_l$ & width of the reference cells in delay \\
\hline
$C_\mathrm{in}$ & \# input channels of the input feature map & $C_\mathrm{out}$ & \# input channels of the output feature map \\
\hline
\end{tabular}}
\label{tab:notation}
\end{table}
\begin{table*}[!htbp]
\centering
\caption{Summary of Channel Estimation Studies in OTFS Systems.}
\footnotesize{\textit{Note: "-" indicates that the corresponding study does not include complexity expression}}
\renewcommand{\arraystretch}{0}
\scriptsize
\begin{tabular}{
|>{\arraybackslash}p{0.5cm}
|>{\arraybackslash}p{0.5cm}
|>{\arraybackslash}p{5.0cm}  % Proposed Method
|>{\arraybackslash}p{1cm}  % Algorithms
|>{\arraybackslash}p{3cm}  % Complexity
|>{\arraybackslash}p{5cm}| % Performance Analysis 
}
\hline
\rowcolor{lightgray}
\textbf{Ref} &
\textbf{Date} &
\textbf{Proposed Method} &
\textbf{Tech.} &
\textbf{Complexity} &
\textbf{Performance Analysis} 
\\ \hline \hline

\cite{8671740} & Mar., 2019& EP-aided CE schemes over multipath channels with integer and fractional Doppler shifts using threshold-based detection & EP & -& BER performance comparison between OFDM across different UE speeds and channel estimation overhead analysis \\ \hline

\cite{8727425} & Aug., 2019 & 3D-structured sparse recovery for downlink CE in massive MIMO systems & 3D-SOMP & - & NMSE and BER comparison between traditional impulse-based techniques and OMP-based techniques \\ \hline

\cite{9110823} & June, 2020& Optimized high-mobility downlink CE scheme for massive MIMO-OTFS networks using fast Bayesian inference and EM framework & Fast EM-VB & $\mathcal{O}(M_A^3 N_{\mathrm{TS}}^3)$& MSE comparison between traditional EM-VB, complexity analysis with traditional EM-VB \\ \hline

\cite{9128497} & June, 2020& OMP and MSP based algorithms for DD-CE in uplink OTFS-MA systems & OMP, MSP & - & NMSE and BER comparison between proposed CS-based and impulse-based CE methods \\ \hline

\cite{9184852} & Sept., 2020 & SBL-based CE algorithm for OTFS modulation & SBL & $\mathcal{O}(N_\mathrm{ite}PN_P^2)$ & Pilot overhead and NMSE comparison with guard pilot-based and OMP \\ \hline

\cite{9200896} & Sept., 2020 & Tensor-based OMP-CE algorithm for uplink transmission of mmWave massive MIMO-OTFS systems with effective pilot design & Tensor-based OMP & $\mathcal{O}(PN_FN_TM_A(M + N + G))$ & Complexity, NMSE comparison between traditional OMP and SOMP techniques \\ \hline

\cite{9303350} & Dec., 2020 & DD domain EP-based CE for mobile OTFS considering residual frame timing offset, carrier frequency offset and fractional multiple Doppler & EP & $\mathcal{O}(MN \log(MN))$& MMSE, BER comparison across different synchronization scenarios\\ \hline

\cite{9400868} & Apr., 2021& Transform-domain basis functions for low-dimensional subspace channel modeling in OTFS with CIPP and LMMSE estimation & CIPP-LMMSE & $\mathcal{O}(4(N_{\mathrm{DO}} + G_{\mathrm{DO}}) N_{\mathrm{DE}} + 2 G_{\mathrm{DE}} N_{\mathrm{DO}})$& MSE and BER comparison with SI-based and CS-based CE methods\\ \hline

\cite{9440710} & May, 2021& Modified sensing matrix based CE for downlink massive MIMO-OTFS systems with fractional Doppler and deterministic pilot design & MSMCE & $\mathcal{O}(P (G_{\mathrm{DE}} + 2 G_{\mathrm{DO}} - 1))$
& NMSE comparison with initial sensing matrix based CE algorithm across different pilot overhead ratios, user velocities, and SNR conditions \\ \hline

\cite{9456029} & June, 2021& MEP algorithm for OTFS-CE with fractional Doppler shifts using Bayesian inference and factor graph representation & MEP & - & NMSE and PAPR comparison with threshold-based methods, CRLB analysis for both bi-orthogonal and rectangular waveforms \\ \hline

\cite{9456894} & June, 2021& Data-aided CE algorithm for SP and data transmission in OTFS systems with iterative interference cancellation & SPA & - & NMSE and BER comparison with coarse estimation, classic approaches, spectral efficiency analysis through superimposed transmission \\ \hline

\cite{9483694} & July, 2021 & Efficient sparse CSI estimation scheme for OTFS systems using BL framework with TF domain pilot transmission & BL, M-BL & $\mathcal{O}(M_\tau^3 N_\nu^3)$ & NMSE and SER comparison between conventional MMSE-based scheme as well as other existing sparse estimation methods \\ \hline

\cite{9539066} & Sept., 2021& SP-based CE and SD framework MEP & SP-I, SP-NI, MEP & SP-NI: $\mathcal{O}(MN) + \mathcal{O}(N_\mathrm{ite} MN P \mathcal{M})$,
SP-I: $\mathcal{O}(MN(N_{\mathrm{out}}+1)) + \mathcal{O}(N_\text{ite} MN P \mathcal{M} (N_{\mathrm{out}}+1))$
 & BER and spectral efficiency comparison with EP schemes across different power allocation ratios, velocity scenarios, and modulation orders \\ \hline
\cite{9590508} & Jan., 2022 & Sparse CSI estimation model for reducing pilot overhead in MIMO-OTFS systems with fractional Doppler handling and row-group sparsity exploitation& RG-OMP, RG-BL & - & NMSE and SER comparison between proposed BL-based schemes and OMP techniques as well as SOTA sparse CE methods \\ \hline
\cite{9686700} & Jan., 2022 & CE method for MIMO-OTFS systems using block SBL with block reorganization for multiple DD path clusters & BSBL-BR & \scalebox{0.9}{$\mathcal{O}(M_{\mathrm{RA}} N_{\mathrm{ite}} N_P (N_{\mathrm{TA}} C_L)^2)$}
 & NMSE and BER comparison between \cite{8671740},\cite{9184852},\cite{8570860},\cite{7140825} \\ \hline
\cite{9685856} & Feb., 2022 & CS-CE scheme for OTFS channels with sparse multipath using continuous DD dictionary without integer delay assumptions& OMPBR & - & Complexity analysis depending on the runtime and NMSE comparison between traditional OMP technique  \\ \hline
\cite{9737331} & Mar., 2022& CE scheme with leakage suppression using smoothness regularization in TF domain & CMDE & - & NMSE and BER comparison with perfect channel matrix denoising estimation (CMDE) across different smoothness regularization weights in high-mobility vehicular scenarios \\ \hline 
\cite{9738478} & Mar., 2022& Off-grid CE scheme using SBL with virtual sampling grid & Off-grid SBL & $\mathcal{O}(5M_\tau^2 N_\nu N_{\mathrm{DE}}N_{\mathrm{DO}})$ for 1D, $\mathcal{O}(5M_\tau^2 N_\nu N_{\mathrm{DE}})$ for 2D & NMSE and BER comparison with on-grid OMP, off-grid Newtonized-OMP (NOMP), and traditional impulse methods across different pilot lengths and SNR conditions \\ \hline
\cite{9771955} & May, 2022 & CE scheme for massive MIMO-OTFS systems exploiting 2D cluster structure in Doppler-angle domain using SBL framework & SBL, LBP & - & NMSE comparison between EM-VB \\ \hline
\cite{9785832} & May, 2022& Joint CE and SD algorithm for OTFS-based LEO satellite communications using VBI & VBI & $\mathcal{O}((M^2 P^3) N_{\mathrm{ite}})$ for CE& NMSE and BER comparison with impulse-based estimation, OMP, pilot-based EM methods across different waveform types and SNR conditions \\ \hline

\cite{9794710} & June, 2022& CE technique for OTFS-SCMA based on CSC using EP-aided sparse-pilot structure & CSC & $\mathcal{O}(M_{\mathrm{dom}} N (J(2k_v + 1) + J^2)N_{\mathrm{ite}})$& BER and spectral efficiency comparison with MSP, QAM-pilot methods across different SNR conditions and user velocities \\ \hline
\cite{9827947} & July, 2022& OTFS-based mmWave MIMO systems with analog and hybrid beamforming and BL-CE &BSBL & - & NMSE comparison between OMP and FOCUSS methods in both analog and hybrid beamforming scenarios \\ \hline

\cite{9864300} & Aug., 2022 & Joint CE and SD for HRIS-aided mmWave-OTFS systems using MEP and EM & JCEDD, MEP, EM & $\mathcal{O}(MN^2 P^2 + \mathcal{M}N^2 P^2)$& NMSE and BER comparison between traditional methods across different SNR conditions and user velocities \\ \hline

\cite{9891774} & Sept., 2022 & CSI estimation model for MIMO-OTFS systems exploiting 4D-sparsity in DDA domain with flexible pilot placement and low training overhead & OMP & - & NMSE comparison between SOTA EP and conventional FOCUSS and MMSE schemes \\ \hline

\cite{9880804} & Sept., 2022 & CE method for fractional Doppler problem using ESBL with Doppler grid segmentation factor & ESBL & $\mathcal{O}\big( (l_{\tau}+1)(2k_{\nu}+2\tilde{N}+ 1)^2 \big)$& NMSE and BER comparison between traditional SBL-based algorithms \\ \hline
\cite{9896637} & Oct., 2022 & Model-driven DL based CE technique for OTFS systems using ResNet to refine preliminary OMP results in DD domain & OMP + ResNet & - & NMSE comparison between traditional OMP technique \\ \hline
\multicolumn{6}{r}{\footnotesize\textit{Continued on the next page}}
\end{tabular}
\label{tab:OranWorkgroups2}
\vspace{-4mm}
\end{table*}

\begin{table*}[!htbp]
\ContinuedFloat
\centering
\caption{{\normalfont\textit{Continued from previous page}.}}
\renewcommand{\arraystretch}{0}
\scriptsize
\begin{tabular}{
|>{\arraybackslash}p{0.6cm}
|>{\arraybackslash}p{0.6cm}
|>{\arraybackslash}p{5.0cm}  % Proposed Method
|>{\arraybackslash}p{1.2cm}  % Algorithms
|>{\arraybackslash}p{3cm}  % Complexity
|>{\arraybackslash}p{5cm}| % Performance Analysis 
}
\hline
\rowcolor{lightgray}
\textbf{Ref} &
\textbf{Date} &
\textbf{Proposed Method} &
\textbf{Tech.} &
\textbf{Complexity} &
\textbf{Performance Analysis} 
\\ \hline \hline
\cite{9928043} & Oct., 2022 & GF-NOMA paradigm with OTFS modulation for LEO satellite-based IoT networks using successive ATI, CE, and SD methods & SOMP, LS & $\mathcal{O}(2G_T(N+1)M_{\mathrm{RA}} N_{\mathrm{IoT}}C_L + N_{\mathrm{IoT}}C_L(N+1)M_{\mathrm{RA}} + 2N_{\mathrm{ite}}^2G_T + N_{\mathrm{ite}}^3 + N_{\mathrm{ite}}G_T(N+1)M_{\mathrm{RA}})$ for SOMP, $\mathcal{O}(
M_{RA} [
(\sum_{k \in N_{\text{act}}} 1 + P_k )^2 G_T
+ (\sum_{k \in N_{\text{act}}} 1 + P_k )^3
+ 2 ( \sum_{k \in N_{\text{act}}} 1 + P_k )
]
)
$ for LS & NMSE and BER comparison between OFDM-based benchmarks across different pilot lengths, transmit powers, and channel conditions \\ \hline

\cite{Wang_2023} &Oct., 2022& CE scheme for doubly selective channels with fractional Doppler and delay using 2D off-grid decomposition and SBL & SBL & $\mathcal{O}(16M^3N^3)$ for virtual sampling rate = 0.5 & NMSE comparison traditional OTFS, 2D off-grid, and doubly fractional methods across different channel conditions \\ \hline

\cite{10001420} & Jan., 2023 & Uplink-aided downlink CE scheme for massive MIMO-OTFS systems analyzing reciprocity and processing latency effects & OMP, LS & - & Latency analysis of uplink and downlink CE depending on processing delay \\ \hline

\cite{10038838} & Feb., 2023 & Joint time and DD domain CE for OTFS systems using SBL to avoid inter-Doppler interference problem & SBL-based BCE & $N_{\mathrm{ite}}[(N_BC_L)^2MN + (N_BC_L)^3 + (N_BC_L)^2 + 2(N_BC_L)MN + MN + N_BC_L]$ & BER and NMSE comparison between EP, LS-based BCE, and perfect CE across different gamma values \\
\hline

\cite{Liu_2023} & Feb., 2023& Low PAPR CE scheme using multiple scattered pilot pattern and SP & SP, MP & - & PAPR and NMSE performance comparison between single pilot schemes \\ \hline

\cite{Zhang_2023} & Feb., 2023& Structured sparsity-based GAMEP algorithm for sparse CE in OTFS underwater acoustic systems & GAMEP, EM & $\mathcal{O}(N_P C_L (7 N_{\mathrm{ite}} + N_{\mathrm{out}}) + N_P (16 N_{\mathrm{ite}} + N_{\mathrm{out}}) + 20 C_L N_{\mathrm{out}})$ & NMSE and BER comparison between traditional impulse-based threshold, conventional AMP, and SBL methods \\ \hline
\cite{10073950} & Mar., 2023 & Low complexity fast greedy sparse recovery algorithm using TCHHTP for DD-CE & TCHTP & 
$\mathcal{O}(N_{\mathrm{out}}N_{\mathrm{ite}}(P^3+MNP^2+MNP))$ & 
NMSE, BER, estimated sparsity, execution time and computational complexity comparison between MSP, OMP, EP techniques \\ \hline

\cite{Muppaneni_2023} & Mar., 2023 & DDIPIC algorithm for OTFS-ISAC systems with joint CE and radar parameter estimation & DDIPIC & $\mathcal{O}\big( 2PMN + 2P^2 MN + P^2 + P^3\big)$ & NMSE and BER comparison between M-MLE with inter-path interference cancellation capability in radar sensing applications \\ \hline

\cite{10100890} & Apr., 2023 & Threshold-based CE for mmWave OTFS systems with nonlinear PA distortions using reformulated pilot signals & Threshold based CE & $\mathcal{O}\big( MN + MNP + N_{\mathrm{ite}} \left( MN^2 + [\log(n_b)]^2 \right) \big)$
 & BER comparison across different SNR conditions, particle sizes, pilot patterns, user velocities, and receive antenna configurations \\ \hline

\cite{RISOTFS21} & Apr., 2023& Uplink transmission scheme and CE design for RIS-aided OTFS systems with efficient sensing and beamforming & EKF& - & NMSE comparison for Doppler estimation, cascaded channel coefficients, and user localization tracking across single-path and multipath scenarios with achievable rate analysis for different RIS beamforming methods \\ \hline

\cite{He2023ATC} & Apr., 2023& Two-stage algorithm for estimating fractional Doppler channels using rough position estimation and quasi-Newton method & Quasi-Newton, Correlation-based & \makecell[l]{
$\mathcal{O}(PDMN +\log N\log(\frac{1}{\epsilon})$ \\
$\cdot \sum_{l=0}^{l_{\tau}} 9P_{\mathrm{DE}}^2)$
}& NMSE and BER comparison with correlation-based and threshold-based algorithms in fractional Doppler channels \\ \hline

\cite{10122554} & May, 2023 & Low-complexity CE method for single-antenna systems with practical rectangular pulses and fractional delays using decoupled joint estimation approach & Decoupled iterative estimation & \makecell[l]{
$\mathcal{O}( P M_{\tau} N_{\nu} + $ \\
$P n_{\nu} N_{\nu} + P m_{\tau} M_{\tau} )$
}& NMSE and SER comparison between Impulse, OMP, SBL-based methods and CRLB with lower complexity \\ \hline
\cite{fd_otfs} & May, 2023 & Two-step OMP with fractional refinement algorithm for SISO radar in OTFS-based ISAC systems handling fractional delays and Doppler shifts & OMPFR & $O(2N_{\mathrm{ite}}N_{\mathrm{dic}}  MN)$& Range and velocity RMSE comparison with fractional refinement versus without fractional refinement, complexity analysis with OMP \\ \hline
\cite{10138432} & May, 2023& DNN-based fractional Doppler CE scheme in air-to-ground communication & DNN-based estimator & $\mathcal{O}(2(l_\tau+1)Q_1 + Q_1Q_2 + 2(1+3P)Q_2 + 2PMN)$& NMSE and BER comparison between conventional methods, complexity analysis with OMP \\ \hline

\cite{10138088} & May, 2023& AP-SP based architecture for CP-aided SISO and MIMO OTFS systems with affine precoding and superimposed pilots & PA-BL, DA-BL, EM & $\mathcal{O}(N_{\mathrm{ite}} (M_\tau N_\nu)^3)$ for SISO, $\mathcal{O}(N_{\mathrm{ite}} (M_\tau N_\nu M_\mathrm{RE})^3)$ for MIMO  & NMSE and SER comparison with FOCUSS, OMP, conventional MMSE, and BCRLB across different SNR conditions, mobility scenarios, and modulation orders\\ \hline

\cite{10147252} & June, 2023& Joint CE and turbo equalization for coded OTFS and OFDM systems using MP and soft MMSE equalizers & BEM, MP, THR-CSI & $\mathcal{O}(MN P \mathcal{M} N_{\mathrm{ite}})$ for MP, $\mathcal{O}(MN P N_{\mathrm{ite}})$ for MMSE & NMSE and BER comparison between coded OTFS and OFDM systems with BPSK and QPSK when BEM-based CE is utilized \\ \hline

\cite{10164149} & June, 2023& Cross-domain CE scheme for OTFS-based mmWave hybrid beamforming systems with off-grid parameters & SBL, EM & $\mathcal{O}(N_{\mathrm{ite}}(N_\alpha^3 + N_\beta^3 + M_\tau^3 + N_\nu^3))$ & NMSE comparison with on-grid and off-grid algorithms \\ \hline

\cite{10151793} & June, 2023& Three overlapped pilot schemes for TF domain OTFS-CE with fractional DD parameters using SIC-enhanced MMSE equalizers & SIC, MMSE, Root-MUSIC & $\mathcal{O}(M_{\mathrm{RA}}[2MN\log_2 MN + N_P^2 M_0 + N_P^3 + \Theta(MN, MN)])$ & NMSE and BER comparison between overlapped pilot designs in both TF and DD domains \\ \hline

\cite{10201832} & Aug., 2023& CS-based CE in MIMO-OTFS systems exploiting structured sparsity with RBOMP & RBOMP & $\mathcal{O}(M_\tau^3N_\nu^3N M_{\mathrm{TA}}^3 + M_\tau^3N_\nu^3 M_{\mathrm{TA}} M_{\mathrm{RA}} N_{\text{P}} M)$
& NMSE comparison between LS, MMSE, OMP, SOMP and existing CS-based techniques and complexity analysis \\ \hline

\cite{10213984} & Aug., 2023& CE for massive MIMO-OTFS systems using 3D-IPRDSOMP to reduce inter-symbol interference. & 3D IPRDSOMP & - & NMSE comparison between traditional LS, OMP, and 3D-SOMP algorithms across different SNR values, pilot overhead ratios, BS antennas, and user velocities \\ \hline

\cite{10209369} & Aug., 2023& $l_0$-norm constrained maximum Versoria criterion based CE and robust MP-MVC symbol detection for OTFS systems under impulsive noise & $l_0$-MVC, MP-MVC &$\mathcal{O}(MN)$& MSE and BER comparison with LMS, maximum correntropy criterion (MCC), and conventional MEP in both Gaussian mixture and $\alpha$-stable noise environments \\ \hline
\cite{10223427} & Aug., 2023& Three OTFS transmission variants with 1D CE: FD-PA-OTFS, TD-PA-OTFS, and TD-TS-OTFS for fast fading channels & FD-PA, TD-PA, TD-TS & $\mathcal{O}((2k_{\tau} + 1)(l_{\nu} + 1)\log(l_{\nu} + 1))$ for FD-PA-OTFS & MSE and BER comparison between variants\\ \hline
\multicolumn{6}{r}{\footnotesize\textit{Continued on the next page}}
\end{tabular}
\vspace{-4mm}
\end{table*}

\begin{table*}[!htbp]
\ContinuedFloat
\centering
\caption{{\normalfont\textit{Continued from previous page}.}}
\renewcommand{\arraystretch}{0}
\scriptsize
\begin{tabular}{
|>{\arraybackslash}p{0.6cm}
|>{\arraybackslash}p{0.6cm}
|>{\arraybackslash}p{5.0cm}  % Proposed Method
|>{\arraybackslash}p{1.2cm}  % Algorithms
|>{\arraybackslash}p{3cm}  % Complexity
|>{\arraybackslash}p{5cm}| % Performance Analysis 
}
\hline
\rowcolor{lightgray}
\textbf{Ref} &
\textbf{Date} &
\textbf{Proposed Method} &
\textbf{Tech.} &
\textbf{Complexity} &
\textbf{Performance Analysis} 
\\ \hline \hline
\cite{10264119} & Sept., 2023& MF-OAMP-based joint CE and data detection for OTFS systems with SP-data placement & OAMP, MF & \shortstack[l]{
\scalebox{0.9}{$\mathcal{O}( N_{\mathrm{out}} (N_{\mathrm{ite}} ( MN \log MN$} \\
\scalebox{0.9}{$+ MN)+ 2MN ))$}
} for CE& NMSE and frame error rate (FER) comparison with threshold-based methods, AMP, and UAMP-SBL \\ \hline

\cite{10272673} & Oct., 2023& Two-stage CE scheme for asymmetrical massive MIMO-OTFS systems with coprime patterns and virtual array reconstruction &  proximal gradient (PG)-OMP & - & NMSE and BER comparison with 3D-OMP, alternating direction method of multipliers (ADMM), and off-grid NOMP\\ \hline
\cite{10292645} & Oct., 2023& VBL-based CE for multi-user OTFS systems exploiting shared sparsity with Gaussian mixture prior & VBL, EM & $\mathcal{O}(M^3 N^3 J^2)$& NMSE and BER comparison with MSP and SBL by exploiting multi-user shared sparsity structure \\ \hline

\cite{10310071} & Nov., 2023& HMM-based CE for OTFS systems with fractional Doppler using UAMEP & UAMEP-HMM, EM & $\mathcal{O}(N_{\mathrm{ite}}(m_\tau+l_\tau)(n_\nu+2k_\nu+\hat{N})C_L)$ & NMSE and BER with UAMEP-SBL, OMP, SBL, and threshold-based methods for fractional Doppler scenarios through structured sparsity exploitation and state evolution analysis \\ \hline

\cite{10356117} & Dec., 2023& Online RG sparse BL-based (ORGBL) CSI estimation for SISO and MIMO-OTFS systems with sparse CE and low-complexity detection & OSBL, ORGBL, EM & $\mathcal{O}(M_\tau^3 N_\nu^3)$ for integer scenario & NMSE and BER comparison with SBL, OMP, FOCUSS, and MMSE \\ \hline

\cite{10370744} & Dec., 2023& 2D pilots for multi-antenna OTFS-CE using perfect sequence based Kronecker array (PKA) with matched filters and code division multiplexing & PKA & $\mathcal{O}(N_\nu^2M_\tau + M_\tau^2N_\nu)$ for inverse-based estimator& NMSE and BER comparison with traditional single pulse pilots and sequence pilots multi-antenna scenarios \\ \hline

\cite{10385066} & Jan., 2024 & Low-complexity DD CE using DZT with EP frame and fractional DD handling & DZT-based decomposition & $\mathcal{O}(3 + M_\tau N_\nu + \frac{M_\tau N_\nu}{2} \log_2(M_\tau N_\nu)(2 + \frac{1}{\Delta\alpha} + \frac{1}{\Delta\beta}) + M_\tau^2 N_\nu^2(\frac{1}{\Delta\alpha} + \frac{1}{\Delta\beta}))$ for Doppler estimation & NMSE and BER comparison with SBL-based methods across different SNR conditions and mobility scenarios \\ \hline

\cite{RISOTFS14} & Jan., 2024 & RIS-assisted OTFS with joint frame structure design, location-aided CE, and IIC detector using delay and shifted-Doppler
domain (DsD) estimator &DsD-based CE & $\mathcal{O}(N_{\mathrm{ite}} l_\tau N^2 P )$ for IIC detector & Symbol error probability (SEP) and NMSE comparison with conventional detectors (MRC, LMMSE, MP) and benchmarks with fractional Doppler effects \\ \hline

\cite{10400942} & Jan., 2024 & Learned denoising-based sparse adaptive CE for underwater acoustic OTFS systems using FastDVDNet with symbol-wise adaptive method & IPNLMS, FastDVDNet, Sparse Adaptive & $\mathcal{O}\big( (M_\tau - l_\tau)\,(N_\nu - 2k_\nu)\,  \big)$ for IPNLMS, $\mathcal{O}\big( (M_\tau - l_\tau)\,(N_\nu - 2k_\nu)\, P \big)$ for denoising & NMSE and BER comparison with OMP and other IPNLMS-based methods\\ \hline

\cite{10409528} & Jan., 2024& Sensing-aided uplink transmission for ISAC vehicular networks with joint parameter association, CE and signal detection using Bi-UAMEP algorithm & PACESD, Bi-UAMEP, MEP & $\mathcal{O}(N_\mathrm{ite}NM^2P \log(MP))$ for SVD preprocessing, $\mathcal{O}(N_\mathrm{ite}(MP + \mathcal{M}))$ & BER and NMSE comparison with MMSE, MEP, and UAMEP-SBL methods \\ \hline
 \cite{10436557} & Feb., 2024 & Grid evolution-based off-grid sparse Bayesian inference (GESBI) for OTFS-CE with rectangular pulses using T-GEESBI algorithm & T-GEESBI, GESBI & $\mathcal{O}(N_{\mathrm{out}} ( 2 (l_\tau+1) (2k_\nu+1) M_\tau N_\nu + N_{\mathrm{ite}} [ \frac{2}{3} ((l_\tau+1)(2k_\nu+1))^3 + 2 M_\tau N_\nu ((l_\tau+1)(2k_\nu+1))^2 + (M_\tau N_\nu)^2 + 4M_\tau N_\nu + 2P(P+1) ] ))$
 & NMSE comparison with other grid evolution-based schemes \\ \hline

\cite{10439989} & Feb., 2024& Learning-based approach for fractional DD CE in OTFS systems using TF domain training with DNNs and computation of a constituent DD parameter matrix (CDDPM) & CDDPM, DNN & - & NMSE and BER comparison with DDIPIC and modified-MLE methods and runtime computational complexity analysis \\ \hline
\cite{10472131} & Mar., 2024 & Joint sparsity pattern learning (JSPL) for massive MIMO-OTFS-CE using Bayesian framework with flexible spike and slab prior model & JSPL, BL, Spike-slab prior & $\mathcal{O}(\frac{P k_\nu}{10} N^2 M^2 N_{\mathrm{TA}} + N_{\mathrm{ite}} N^2 M^2 N_{\mathrm{TA}})$ & NMSE and BER comparison with impulse-based, OMP, and 3D-SOMP methods \\ \hline

\cite{10475591} & Mar., 2024 & Joint CE and SD using SP in OTFS systems with BEM and MEP receiver & SP-DD, BEM, MEP &\scalebox{0.9}{$\mathcal{O}(Q_\mathrm{BEM}MN\log(MN)N_{\mathrm{out}})$}& BER and NMSE comparison with SP-BEM-MEP and SP-aided-MEP methods \\ \hline

\cite{10475894} & Mar., 2024 & SBL-based off-grid estimation for OTFS channels with Doppler squint effect using EP pattern & SBL-DSE, Off-grid SBL & $\mathcal{O}(N_{\mathrm{ite}} \cdot M_\tau \cdot N_\nu\cdot (l_\tau (N_\beta+1))^2)$& NMSE comparison with OMP-DSE and OMP-NDSE methods considering Doppler squint effect \\ \hline

\cite{10476726} & Mar., 2024 & Joint CE and SD for MIMO-OTFS systems using ABOMP and joint detection algorithm with partial iterative feedback & ABOMP, JDCE, AMP & $\mathcal{O}(N_{\mathrm{ite}} \sum_{j=1}^{N_{\mathrm{ite}}} |\Psi^j| N_e M_{\mathrm{TA}}$
$\times\mathcal{M})$ & NMSE and BER comparison with OMP, BOMP, SBL-EM, and BSBL-BR methods \\ \hline

\cite{Priya_2024} & Apr., 2024 & Two-stage CE for OTFS in overspread channels using DD domain EP and time domain dual chirp correlation with modified MRC detection & Two-stage CE, Dual chirp, MRC & $\mathcal{O}(MN(\log_2 N + |\mathcal{B}||\mathcal{K_{\hat{\ell}}}||\mathcal{J_{\hat{\ell}}}| + l_{\tau}))$ & BER and NMSE comparison with MP methods across different channel models \\ \hline

\cite{Yang_2024} & Apr., 2024 & IPPS and inter-path interference mitigation for fractional DD-CE using L-BFGS method & IPPS, L-BFGS& $\mathcal{O}\Big( N_\mathrm{ite} \big( \lambda m_\tau M_\tau^2 N_\nu + \lambda n_\nu N_\nu^2 M_\tau + \lambda^2 M_\tau^2 N_\nu^2 \big) \Big)$  for improved IPPS & NMSE and BER comparison with low-complexity CE methods \\ \hline

\cite{10506450} & Apr., 2024 & Off-grid fractional DD-CE using Bayesian compressive sensing with LSM prior for OTFS systems & OG-BCS, FBCS & \parbox{4cm}{
\scalebox{0.9}{$\mathcal{O} \bigg[ 12\, \Bigg( \frac{M_p N_p}{\log(M_\tau N_v)} \Bigg)^2 M_p N_p +$} \\
\scalebox{0.9}{$(M_p N_p)^2 M_p N_p +$} \\
\scalebox{0.9}{$(M_p N_p)^2 \Bigg( \frac{M_p N_p}{\log(M_\tau N_v)} + 1 \Bigg) \bigg]$}
}& NMSE and runtime performance comparisons between OG-FBCS , SBL, Laplace and OMP under fractional DD settings with varying paths, pilots, and DD grid configurations \\ \hline

\cite{10540188} & May, 2024 & Nonlinear HPA-aware CE for OTFS systems using LS-based LSTM NNs in TF domain & LS, LSTM-NN & $\mathcal{O}(M_{\mathrm{on}}^2 + N_P^2 + M_{\mathrm{on}} N_P + M_d N)
$ &BER, PAPR, and throughput analysis in high-mobility and nonlinear HPA scenarios compared to classical DD domain estimation schemes \\ \hline
\cite{10550432} & June, 2024 & Two-stage iterative CE using HSP design with center and edge pilots & TSICE &- & BER and PAPR comparison between SSIP and SSPP methods \\ \hline

\cite{10557727} & June, 2024 & WLS channel parameter estimation algorithm for OTFS systems with fractional Doppler shifts & WLS, IIS & $\mathcal{O}\big(N_{\text{ite}}\, (N \log N + P^2)\big)$& NMSE comparison with CRLB bounds, NOMP, ML, and LS algorithms \\ \hline
\cite{Zhang_2024} & June, 2024 & Off-grid SBL for OTFS-CE with fractional Doppler shifts using MM algorithm & SBL, MM, EM & $\mathcal{O}\big( N_\mathrm{ite}N_p M_p \big[ ( \lceil k_{\max} \rceil + \lceil k_{\min} \rceil + 1 ) ( l_{\max} + 1 ) \big]^2 \big)
$ & NMSE comparison with on-grid OMP, VBI, and approximate methods in fractional Doppler scenarios \\ \hline

\multicolumn{6}{r}{\footnotesize\textit{Continued on the next page}}
\end{tabular}
\vspace{-4mm}
\end{table*}

\begin{table*}[!htbp]
\ContinuedFloat
\centering
\caption{{\normalfont\textit{Continued from previous page}.}}
\renewcommand{\arraystretch}{0}
\scriptsize
\begin{tabular}{
|>{\arraybackslash}p{0.6cm}
|>{\arraybackslash}p{0.6cm}
|>{\arraybackslash}p{5.0cm}  % Proposed Method
|>{\arraybackslash}p{1.2cm}  % Algorithms
|>{\arraybackslash}p{3cm}  % Complexity
|>{\arraybackslash}p{5cm}| % Performance Analysis 
}
\hline
\rowcolor{lightgray}
\textbf{Ref} &
\textbf{Date} &
\textbf{Proposed Method} &
\textbf{Tech.} &
\textbf{Complexity} &
\textbf{Performance Analysis} 
\\ \hline \hline

\cite{10582886} & July, 2024 & Matrix Pencil-based CE algorithm for OTFS systems with fractional Doppler shifts using IDFT transformation & Matrix Pencil& $\mathcal{O}\big( l_{\tau} N \log N + \sum_{l}^{l + l_{\tau}} [ P_l^3 + ((N+1)/2)^3 \log (1/\chi) ] \big)$& NMSE and BER comparison with threshold, correlation, SBL, and polynomial methods \\ \hline

\cite{10587305} & July, 2024 & User-specific CE overhead optimization and resource allocation scheme for multi-user OTFS systems with IGI mitigation & overhead optimization &- & CDF and capacity performance comparison across various SNR conditions and bandwidth allocations\\ \hline

\cite{Guo_2024} & July, 2024 & CS-BEM OTFS-CE with dense pilot insertion and COMP algorithm & CS-BEM, COMP & $\mathcal{O}\bigg[
N^2 M + (2 M_\mathrm{dom} C_L + 3 M_\mathrm{dom}^2 + C_L) N_P Q_\mathrm{BEM} +
\big( M^2 N^2 C_L Q_\mathrm{BEM} + (C_L^2 Q_\mathrm{BEM}^2 + C_L Q_\mathrm{BEM}) M N + C_L^3 Q_\mathrm{BEM}^3 \big) (N_\mathrm{out} - 1)
\bigg]
$& BER and NMSE comparison with CS-COMP, BEM receiver, and EP methods \\ \hline

\cite{10612829} & July, 2024 & SSAMP algorithm for MIMO-OTFS-CE without prior sparsity knowledge & SSAMP, MSSAMP & - & NMSE comparison with OMP (known/unknown K), SOMP, and various SSAMP configurations \\ \hline
\cite{10614842} & July, 2024 & Zero-Bin Low-PAPR pilot design for Zero-Pad OTFS systems with two-step CE method & ZBLP, MRC & - & NMSE and BER comparison with EP methods, PAPR analysis for different $M$ and $N$ values, and pilot overhead comparison \\ \hline

\cite{10637960} & Aug., 2024 & CE scheme under data interference for EP-based OTFS-CE using non-rectangular windowing and CRLB minimization & Window- ing, PSO optimization & - & NMSE and BER comparison under varying guard intervals, Doppler shifts, and data interference levels; CRLB analysis across rectangular, Kaiser, and Slepian windows with optimized pilot sequences \\ \hline

\cite{10654761} & Aug., 2024 & DRSNN with sparse prior for denoising-based OTFS-CE & LS, DRSNN & $\mathcal{O}((l_\tau+1)(2k_\nu+2P+1) \log (l_\tau+1)(2k_\nu+2P+1) + N_{\mathrm{ite}}L_n \sum_{i}^{L_\mathrm{DB}} c_{i-1}s_i^2c_i + (l_\tau+1)(2k_\nu+2P+1) \sum_{l} c_{l-1}s_l^2c_l)$ & NMSE and failure rate comparison with LMMSE, SBL, DCNN under AWGN and t-distribution noise\\ \hline

\cite{10630836} & Aug., 2024 & SBL and block-sparse learning for OTFS-based mmWave MIMO-ISAC systems enabling joint radar target and CE & BL, B-BL & $\mathcal{O}((M_\tau N_\nu)^3)$ for BL and $\mathcal{O}((M_\tau N_\nu N_{\mathrm{RF}})^3)$ for B-BL & NMSE and SER evaluations in phased array-ISAC and mmWave MIMO-ISAC configurations\\ \hline

\cite{10640141} & Aug., 2024 & Multiple-frame-based coupled SBL for SP OTFS-CE & M-CPSBL, FM-CPSBL & $\mathcal{O}((M_T N_\nu)^3 + (2MN+3)(M_T N_\nu)^2)$ & NMSE, BER, and SE comparison between MMSE-SP and EP methods \\ \hline

\cite{10677422} & Sept., 2024 & 2D off-grid CE for OTFS using equidistributed grid evolution guided by estimated path distribution & Grid evolution, SBL & $\mathcal{O}(N_\alpha^2 N N_\beta)$ & NMSE and BER comparison between uniform 2D/1D off-grid and 2D-SBL methods\\ \hline

\cite{10683792} & Sept., 2024 & Joint channel and data estimation algorithm via parametric bilinear Gaussian belief propagation for OTFS demodulation in doubly-selective fading channels & PBiGaBP, JCDE, GaBP & $\mathcal{O}(N_\mathrm{ite}MNP)$& BER and NMSE comparison with MEP-based methods and genie-aided schemes \\ \hline

\cite{10666708} & Sept., 2024 & ST-based CE method for OTFS systems using sequential parameter estimation with threshold-based detection & ST-CE, Threshold-based detection & $\mathcal{O}(PMN)$ & NMSE and BER comparison with EP-CE, LIB-CE, and SIB-CE methods \\ \hline

\cite{10690180} & Sept., 2024 & Adaptive block sparse backtracking (ABSB) for enhanced CE in massive MIMO-OTFS systems & ABSB & $\mathcal{O}(N_\mathrm{ite}[K_\mathrm{SE}N+(K_\mathrm{SE}/\mathcal{B})\log(K_\mathrm{SE}/\mathcal{B})+(24)\mathcal{B}^3+\mathcal{B}N])$ & NMSE comparison with OMP, BOMP, CSMP, and ROMP methods \\ \hline

\cite{10700683} &Sept.,  2024 & Impulse-based CE method for OTFS systems addressing fractional delay and Doppler shifts using low-complexity algorithm with pulsone approach &MMSE& $\mathcal{O}(\alpha_{\max}P + 1)MN\log(MN^2) + \alpha_{\max}P\log(MN) + P)$ & NMSE and BER comparison with CRLB bounds \\ \hline

\cite{10704024} &Oct.,  2024 & Grid evolution (GE) method for doubly fractional CE using adaptive grid refinement with fission and adjustment processes & GE, SBL& - & NMSE and computational complexity comparison with on-grid, 1D off-grid, and 2D off-grid methods \\ \hline
\cite{10711236} & Oct., 2024 & Adaptive pattern-coupled SBL (APCSBL) algorithm for CE using hierarchical Gaussian prior model & APCSBL, Gaussian prior model, EM & $\mathcal{O}(N_{\mathrm{ite}}(N_\nu^3 + N_\nu^2 + M_\tau N_\nu))$ & NMSE and BER comparison with EP, OMP, SBL, PCSBL methods \\ \hline

\cite{10714398} & Oct., 2024 & OTFS-based ISAC framework for environment sensing with target state and transmitter location estimation & WLS, SDR & - & MSE comparison for position and velocity estimation \\ \hline

\cite{10742117} & Nov., 2024 & Pilot-aided joint time synchronization and CE (JTSCE) algorithm for OTFS systems using MLS with timing offset estimation and off-grid Doppler estimation & JTSCE, MLS, Off-grid estimation & - & BER comparison with perfect synchronization and CSI scenarios \\ \hline

\cite{10747184} & Nov., 2024 & DL-based 3-phase channel prediction scheme for high-mobility MIMO-OTFS systems using autoencoder and RNN & 3-phase prediction, RNN & $\mathcal{O}(l_\tau k_{\nu} M_A)$ & BER and NMSE comparison with traditional pilot structures and path prediction methods \\ \hline

\cite{10752433} & Nov., 2024 & SFPCE and de-interference iterative detection (DIID) algorithms for multi-user OTFS-CE uplink & SFPCE, MMSE, MP, DIID, HSP & $\mathcal{O} ( (P^2MN + P^3)(J + JN_{\mathrm{ite}}) )$& BER and MSE comparison with SSPCE, SSIPCE \\ \hline
\cite{10758799} & Nov., 2024 & VBL-based sparse CE for quantized MIMO-OTFS/OTSM with finite-resolution ADCs & VBL, EM-SBL, MMSE& $\mathcal{O}(M_\tau^2N_\nu^2N_\mathrm{TA}^2MN_P)$ & NMSE and SER analysis under various ADC resolutions; comparison with OMP, FOCUSS, and MMSE methods \\ \hline

\cite{10815946} & Dec., 2024 & GAN-based CE for OTFS-UAV systems in high-mobility scenarios & GAN, U-Net, PatchGAN& $\mathcal{O}(N_\mathrm{ite}  (B_\mathrm{ED} + L_L)  (M N P  s^2  N_\mathrm{conv}))$ & BER, OP, and NMSE analysis for various velocities and modulation schemes; complexity and processing time comparison with baseline estimators \\ \hline

\cite{10816508} & Dec., 2024 & Differential modulation-aided CE with lightweight enhanced CE network (LEn-CENet) assistance for uplink OTFS systems & decision- feedback, LEn-CENet & $\mathcal{O}(24M^2N^2 +3 M^2N + N^2M +4 MN)$& NMSE and BER analysis under various SNRs, velocities, and Doppler bins; comparison with classic pilot-aided and SP-aided CE methods \\ \hline

\multicolumn{6}{r}{\footnotesize\textit{Continued on the next page}}
\end{tabular}
\end{table*}

\begin{table*}[!htbp]
\ContinuedFloat
\centering
\caption{{\normalfont\textit{Continued from previous page}.}}
\renewcommand{\arraystretch}{0}
\scriptsize
\begin{tabular}{
|>{\arraybackslash}p{0.6cm}
|>{\arraybackslash}p{0.6cm}
|>{\arraybackslash}p{5.0cm}  % Proposed Method
|>{\arraybackslash}p{1.2cm}  % Algorithms
|>{\arraybackslash}p{3cm}  % Complexity
|>{\arraybackslash}p{5cm}| % Performance Analysis 
}
\hline
\rowcolor{lightgray}
\textbf{Ref} &
\textbf{Date} &
\textbf{Proposed Method} &
\textbf{Tech.} &
\textbf{Complexity} &
\textbf{Performance Analysis} 
\\ \hline \hline
\cite{10856399} & Jan., 2025 & DRAN for pilot-data interference mitigation in OTFS-CE & DL, DRAN, filtering & - & NMSE comparison with classical and learning-based OTFS-CE methods under various channel conditions \\ \hline

\cite{10870378} & Feb., 2025 & Block time-varying CE using limited pilots in OTFS & ML, EKF & - &MSE and FLOPs performance analysis versus block-type and BCRLB bounds \\ \hline

\cite{10887291} & Feb., 2025 & Low-complexity pilot-assisted alternative iterative (PAAI)-CE and detection using generalized compressed sensing (GCS)-BEM pilot insertion strategy and PAAI-CE method & GCS-BEM, PAAI-CE& $\mathcal{O}\big( M^2 N^2 P\, \sum_{i=1}^{N_{\mathrm{ite}}} Q_{\mathrm{BEM}}^{i} \big)$ & NMSE and BER comparison with CS-BEM methods \\ \hline

\cite{10916513} & Mar., 2025 & Model-driven CE network with fractional Doppler handling in delay-time domain & CENet, LS &- & MSE and BER comparison with LS, threshold, and DNN-based estimators; computational complexity analysis with FLOPs \\ \hline

\cite{10915571} & Mar., 2025 & Fast sparse recovery for MIMO-OTFS using hybrid burst-sparsity prior and VBI factorization with angle-Doppler refinement & VBI, OGVBI & $\mathcal{O}(N_\mathrm{ite}N_P^2M_{\mathrm{RA}}^2M_{\tau}G)$& NMSE comparison with LS, l1-norm, OGVBI methods \\ \hline

\cite{10937496} & Mar., 2025 & Doubly fractional CE scheme using segmentation-based SBL with adaptive factors and 2D processing using fixed-point update rule (FPUR) and weighted average
fusion (WAF)& S-SBL, WAF, FPUR & $\mathcal{O}(N_{\mathrm{ite}}(N_{\alpha}^2N_\nu + N_{\beta}^2M_\tau))$ & NMSE and BER comparison with SBL, ESBL, and off-grid methods \\ \hline

\cite{10938192} & Mar., 2025 & Smoothed L0 algorithm for OTFS-CE with unknown channel sparsity using CS approach & SL0, CS & $\mathcal{O}(N_\mathrm{ite}4k_{\nu}(l_{\tau}+1))$ & NMSE and BER comparison with SAMP, OMP, and LS methods\\ \hline

\cite{10970022} & Apr., 2025 & CE and hybrid precoding for massive MIMO-OTFS addressing doubly squint effect with chirp pilots & Hybrid precoding & $\mathcal{O}(M_AM\log(M) + M_A + 3M)$ & NMSE comparison across different squint parameters; achievable rate analysis versus SNR, bandwidth, and user velocity \\ \hline

\cite{10976498}& Apr., 2025 & 2D off-grid sparse CE for RIS-aided OTFS using multi-frame block CPSBL (B-CPSBL) with phase optimization and dimension-wise sinusoidal
maximization (DSM) & B-CPSBL, DSM & $\mathcal{O}(N_{\mathrm{ite}}(N_P^3 + N_\nu)^2Q_F^2N_P))$ for B-CPSBL & NMSE and BER comparison with OMP, SBL, and PCSBL methods; achievable rate analysis with phase optimization algorithms \\ \hline

\cite{10993437} & May, 2025 & Pilot-aided CE algorithm for OTFS in high-dynamic scenarios based on MDCFT, enabling estimation of second-order phase in DD domain & MDCFT& $\mathcal{O}((N_a + 1)MN\log_2 N)$ & BER and RMSE comparison with time-domain compensation, conventional pilot/PN-based methods \\ \hline

\cite{10994238} & May, 2025 & Symbiotic radio communication over OTFS with DD-domain amplitude-phase and CF modulation, enabling joint primary/secondary symbol embedding and robust CE & Off-grid SBL, CF modulation, ECSIEst-Net, SSymDet-Net & - & NMSE and BER comparison with LS, PC-SBL, and data-driven estimators; complexity analysis in FLOPs\\ \hline
\cite{11010104} & May, 2025 & DL-based secondary information detection and DL-SBL CE for RIS-aided symbiotic radio OTFS systems; RIS partitioned for secondary transmission with structured frequencies & SecNet, DL-SBL & - & BER and MSE comparison with FFT, ESPRIT, MUSIC, SBL, structured SBL \\ \hline

\cite{11014223} & May, 2025 & Low-complexity BEM-based OTFS-CE under hardware impairments, with theoretical MSE/BER analysis & BEM, MMSE & $\mathcal{O}((Q_\mathrm{BEM} + 1)(l_\tau + 1)^3)$ & Analytical and simulated MSE/BER comparison under different hardware quality factors\\ \hline

\cite{11023080} & June, 2025 & VRCF OTFS-CE systems addressing inter-Doppler interference in fractional Doppler scenarios & VRCF & $\mathcal{O}(l_{\tau}(k_\nu(P + N) + (N_\mathrm{GS} + N_\mathrm{ite} + N_\mathrm{out})(PN + \sum_{l=0}^{l_{\tau}} P_l^2)))$ & NMSE and BER comparison with threshold, cross-correlation, ML, and spline interpolation methods \\ \hline

\cite{11030215} & June, 2025 & Iterative Zadoff-Chu sequence-based CE (ZCE), enabling low-PAPR pilots and robust CE for both integer and fractional Doppler & ZCE, MP& $\mathcal{O}\big( N_{\mathrm{ite}} [ N_{\mathrm{out}} N M T_A \mathcal{M} + N M T_A + N M N_P + Nl_\tau (g_k + r_k)(g_l + r_l) ] \big)$
 & NMSE, BER, and PAPR comparison with EPACE, SBL, ISP, SSP \\ \hline

\cite{11053775} & June, 2025& Parameter-inherited SBL and UAMEP-SBL based CE for OTFS-assisted ISAC; joint DD estimation and SDR validation & PI-SBL, UAMP-SBL, EM-SBL& $\mathcal{O}(MN \log(MN))$ for PI-SBL, $\mathcal{O}((MN)^2)$ for UAMP-SBL, $\mathcal{O}((MN)^3)$ for EM-SBL & Experimental NMSE and BER, rapid convergence, complexity and sensing parameter accuracy comparison \\ \hline

\cite{11058951} & June, 2025& DD pilot structure and robust CE via norm-zero-modified generalized MVC (MGMVC) for OTFS-ISAC systems with using all-delay pilot (ADP) & l$_0$-MGMVC, ADP & - & PAPR, BER, MSD, and range-speed comparison with SIP, SP, OMP, LS, MCC, SRS, MVC estimators \\ \hline
\cite{11068135} & July, 2025 & Low-complexity DD-domain path estimation algorithm for OTFS-ISAC addressing inter-path interference & Sequential estimation& $\mathcal{O}(MN \log(MN) + P_{\max}(m_\tau M + n_\nu N + MN))$ & NMSE, Delay/Doppler/Channel gain MSE, SER comparison with ML and DDIPIC \\ \hline

\cite{11071963} & July, 2025 & DL-based JCEDD for OTFS, using data padding and slicing augmentation techniques & DL-JCEDD & \makecell[l]{
$\mathcal{O}\Bigg(\sum_{l=1}^{N_\mathrm{conv}} \Big(2MNs^2$ \\
$C_{\mathrm{in}} C_{\mathrm{out}} + 3MNC_{\mathrm{out}}$ \\
$+\, MNC_{\mathrm{out}}\Big)\Bigg)$
}&NMSE and BER comparison between conventional JCEDD and pilot-based schemes \\ \hline

\cite{11072043} & July, 2025 & Joint DD domain channel and clipping amplitude estimation, channel tracking, and iterative detection for clipped OTFS systems & SBL-JE, KF, MMSE-DFBE & 
$\mathcal{O}(M_\tau^3 N_\nu^3 + M_\tau^2 N_\nu^2 + MN M_\tau N_\nu + M_\tau N_\nu + \mathcal{B}MN M_\tau N_\nu)$ for SBL-JE, $\mathcal{O}(M^2N^2)$ for KF, $\mathcal{O}(PM^2N^2)$ for MMSE-DFBE & NMSE and BER comparison for joint estimation, tracking, and detection under nonlinear PA effects\\ \hline

\cite{11072479} & July, 2025 & ASGR with fast FBCS for doubly fractional channel estimation in OTFS & ASGR, FBCS & $\mathcal{O}( (M_pN_p+3)(M_\tau N_\nu + P))$ & NMSE comparison with OG-SBL, SBL, OMP, GE-SBL, OG-FBCS; grid refinement impact on estimation accuracy \\ \hline
\multicolumn{6}{r}{\footnotesize\textit{Continued on the next page}}
\end{tabular}
\end{table*}
\bibliographystyle{IEEEtran}
%\bibliography{Referanslar}
\clearpage
\bibliography{A_Referanslar}
\end{document}